\documentclass[twocolumn, prx, superscriptaddress,notitlepage]{revtex4-2}
\usepackage{graphicx}% Include figure files
\usepackage{dcolumn}% Align Table columns on decimal point
\usepackage{bm}% bold math
\usepackage[usenames,dvipsnames]{color}
\usepackage{mathtools}
\usepackage{lineno}
\usepackage{amsmath}
\DeclareMathOperator*{\argmax}{arg\,max}

\usepackage[most]{tcolorbox}
\usepackage{multirow}
\usepackage{gensymb}
\usepackage[normalem]{ulem}
\usepackage{CJK}
\usepackage{comment}
\usepackage[colorlinks, linkcolor=blue,anchorcolor=blue,citecolor=blue,urlcolor=blue]{hyperref}
\usepackage{amssymb}
\usepackage{pifont}

\usepackage{pdfpages}

\makeatletter
\AtBeginDocument{\let\LS@rot\@undefined}
\makeatother

\usepackage{physics}
\usepackage{natbib}
\setcitestyle{super}
\usepackage{booktabs}
\usepackage{colortbl}
\usepackage[multiple]{footmisc}
\usepackage{color,soul}
\usepackage[mathscr]{euscript}
\begin{document}
\begin{CJK*}{UTF8}{}

\title{ 
Quantum counterdiabatic driving enhanced by two-stage local control}

\author{Changhao Li}
\email{changhaoli96@gmail.com}
\thanks{These authors contributed equally to this work.}
\affiliation{Global Technology Applied Research, JPMorgan Chase, New York, NY 10017, USA}
\author{Jiayu Shen}
\email{jiayu.shen@jpmchase.com}
\thanks{These authors contributed equally to this work.}
\affiliation{Global Technology Applied Research, JPMorgan Chase, New York, NY 10017, USA}
\author{Ruslan Shaydulin}
\affiliation{Global Technology Applied Research, JPMorgan Chase, New York, NY 10017, USA}
\author{Marco Pistoia}
\affiliation{Global Technology Applied Research, JPMorgan Chase, New York, NY 10017, USA}

\begin{abstract}

Counterdiabatic driving can suppress diabatic losses during adiabatic ground state preparation. As its implementation necessitates the generation of an adiabatic gauge potential, which 
requires knowledge of the spectral gap of instantaneous Hamiltonians and involves highly non-local controls in many-body systems, local counterdiabatic  driving with approximate adiabatic gauge potential has been widely adopted. In this work, using transverse-field Ising model as an example, we present an in-depth study of the performance of local counterdiabatic protocols and provide a closed-form analysis of ground-state-fidelity optimization.
We then propose a  two-stage protocol based on local counterdiabatic and simple local single-body control to further improve the performance. 
The practical implementation of these protocols does not require diagonalization of instantaneous Hamiltonians or spectral gaps, and only additional local single-body driving is involved.
To benchmark their performance, we experimentally implement digitized adiabatic quantum evolution in a trapped-ion system.
We characterize the quality of the prepared states
and explore the scaling behavior with system size up to 14 qubits. Our demonstration of 
quantum shortcut to adiabaticity opens a path towards preparing ground states of complex systems with accessible local controls.

\end{abstract}

\maketitle
\end{CJK*}	

\section*{Introduction}
Preparation of the ground state of quantum many-body Hamiltonians is a difficult problem with ubiquitous applications which include quantum optimization and quantum optimal control~\cite{crosson2014different,Koch2022,abbas2023quantumOptimization,dalzell2023quantum}. 
A well-established protocol for this problem is to
adiabatically interpolate between a Hamiltonian with an easy-to-prepare ground state and a Hamiltonian with the target ground state~\cite{RevModPhys.90.015002,Barends2016,Aharonov2008}. 
However, the implementation of adiabatic control requires the driving to be slow, leading to long time scales. To minimize losses due to diabatic transitions during the evolution within a fixed total time, a number of approaches known as shortcut to adiabaticity have been proposed~\cite{STA_RevModPhys.91.045001,Sels2017,PRL2019_FE,Kolodrubetz2017,Demirplak2003,Demirplak2005,Passarelli2020PRR} with broad applications in both quantum physics and classical stochastic systems~\cite{NaturePhyIram2020}.

 Among these techniques, 
a common approach to suppress diabatic transitions is counterdiabatic (CD) driving, which adds a velocity-dependent term to the control Hamiltonian to compensate the diabatic losses to excited states during the evolution~\cite{Berry2009}. The CD driving term is constructed using an adiabatic gauge potential that characterizes adiabatic deformations between quantum eigenstates and is closely related to the quantum geometry of eigenstates~\cite{Kolodrubetz2017}.
While it is known to exist in principle, obtaining and implementing this gauge potential in many-body systems is a formidable task. Its realization demands knowledge of the spectral properties of instantaneous Hamiltonians and the manipulation of complex, highly non-local multi-body interactions.
To tackle this challenge, local CD~\cite{Sels2017,Kolodrubetz2017,PhysRevA.90.060301,morawetz2024efficient,COLD_PRXQuantum.4.010312,PhysRevApplied.13.044059,hartmann2019rapid,Michele2023counterdiabatic,keever2023adiabatic,PhysRevA.93.062108,PhysRevResearch.4.043204} (LCD) driving protocols have been proposed to suppress diabatic transitions with readily accessible local controls.
While they can enhance target state preparation within certain parameter regimes, their efficacy and performance in generic models remain unexplored, and the potential for improvement in scenarios where LCD fails to achieve high-fidelity states is still uncertain. 
While first-order LCD for the TFIM is well established, two practical issues remain when the goal is high-fidelity ground-state preparation on near-term hardware. One is that the commonly-used variational method optimizes a spectrum-averaged operator metric and does not necessarily maximize ground-state fidelity, motivating additional optimization and analysis. The second is that single-body LCD misses interaction-generated corrections that appear in higher-order gauge potentials and can become important in certain parameter regimes.

In this work, we  perform an in-depth study of local counterdiabatic driving protocols for ground state preparation across various parameter regimes, using transverse-field Ising model as an example. 
Based on well-established variational method in finding single-body adiabatic gauge potential, we firstly find a closed-form guideline for maximizing ground-state fidelity. 
While effective within specific parameter regimes, fidelity rapidly decreases when the two-body interaction term in the Ising model predominates the target Hamiltonian. 
To achieve consistently high fidelity, motivated by higher-order adiabatic gauge potential and symmetry properties of the model, we then design a two-stage protocol leveraging LCD and simple local controls. 
This straightforward approach significantly improves the ability of the protocol to mitigate diabatic transitions induced by interactions. 
Compared to approaches that optimize time-dependent multi-parameter control fields~\cite{COLD_PRXQuantum.4.010312,Barone2024NJP,morawetz2024efficient}, 
the ground-state-fidelity optimization protocol here is complementary in that it minimizes the number of additional control degrees of freedom during the sweep,
as it  adds only two global scalar parameters (the LCD rescaling $\lambda_f$ and, for LCDLU, a single rotation angle).

To benchmark their performance, we conduct digitized adiabatic quantum evolution experiments in a trapped-ion system.
We assess the fidelity of state preparation through state tomography and energy measurements, investigating scalability in systems up to 14 qubits. This demonstration of a local control-based quantum shortcut to adiabaticity paves the way for preparing the ground state of complex systems using readily available controls and would find applications in various fields including entangled state preparation~\cite{STA_RevModPhys.91.045001}, quantum optimization~\cite{crosson2014different,abbas2023quantumOptimization} and adiabatic quantum computing~\cite{RevModPhys.90.015002}.

\section*{Results}
\subsection{Approximate gauge potential}

A central task in quantum control and adiabatic quantum computing is to transport a stationary state from an initial value of the control parameter $\lambda(0)$ at $t = 0$, to a final value $\lambda(\tau)$ at $t = \tau$. While the standard adiabatic approach requires long time scales, the CD driving utilizes an auxiliary term to 
compensate the diabatic excitations that arise when the parameter $\lambda(t)$ is varied at a finite rate. That is, we have $H_{\mathrm{CD}}(t)  = H(\lambda(t)) + \dot{\lambda} A_{\lambda}$,
where the adiabatic gauge potential (AGP) $A_{\lambda}$~\cite{Kolodrubetz2017} satisfies 
\begin{equation}\label{eq:exact_AGP}
    \bra{m}A_{\lambda}\ket{n} = i\bra{m}\ket{\partial_\lambda n } = -i\frac{\bra{m}\partial_\lambda H \ket{n}}{\epsilon_m - \epsilon_n},
\end{equation}
and $\ket{n}$ and $\epsilon_n$ denote the eigenstates and associated energy of the Hamiltonian $H{(\lambda})$.
As Eq.~\ref{eq:exact_AGP} suggests, finding the exact AGP requires knowing the system's spectral information,
which in general requires the diagonalization of instantaneous Hamiltonian and is intractable for many-body systems. Furthermore, the AGP usually involves non-local multi-body terms. To address the issues,
approximate gauge potential ~\cite{Sels2017,PRL2019_FE,Kolodrubetz2017} has been proposed using, for example, 
nested commutator expansion
\begin{equation}\label{eq:approx_AGP}
    A_{\lambda}^{(l)}  =  i  \sum_{k=1}^{l} \alpha_k\underbrace{[H,[H,\dots [H}_{2k-1}, \partial_{\lambda}H]]].
\end{equation}
The coefficients $\alpha_k$ can be found by minimizing the action $ S(A_{\lambda}^{(l)}) = \tr(G_l^{2})$ with $  G_l(A_{\lambda}^{(l)}) = \partial_{\lambda}H + i [A_{\lambda}^{(l)}, H] $ (see Supplementary Note 1 for details),
and the process does not require diagonalization of instantaneous Hamiltonian. When taking the exact AGP, the $G$ operator is diagonal in the instantaneous eigenbasis and has minimal Hilbert-Schmidt norm~\cite{Sels2017,Kolodrubetz2017}.
We remark that higher orders of the approximate gauge potential would involve more variational coefficients $\alpha_k$, thus leading to a better approximation of the exact AGP. However, this comes at the cost of implementing non-local multi-body terms. The effect of the approximate gauge potential can be found as
\begin{equation}\label{eq:exact_AGP_approximate}
\begin{aligned}
    \bra{m}A_{\lambda}^{(l)}\ket{n} & =i \sum_{k=1}^{l} \alpha_k \bra{m}\underbrace{[H,[H,\dots [H}_{2k-1}, \partial_{\lambda}H]]] \ket{n} \\
    & = i \left(\sum_{k=1}^l \alpha_k (\epsilon_m - \epsilon_n)^{2k-1}\right)\bra{m}\partial_\lambda H \ket{n}.
\end{aligned}
\end{equation}

\subsection{Model}
To investigate the performance of LCD-based driving protocols using approximate gauge potential discussed above, we consider the following model as an example:
\begin{equation}
H(t) = \sum_{j=1}^L \left( h_{z} (t)\sigma_{j}^{z} + h_{x}(t) \sigma_{j}^{x}+ J(t) \sigma_{j}^{z}\sigma_{j+1}^{z} \right)
\label{eq:TFIM_model}
\end{equation}
where $\sigma^{x}, \sigma^z$ are Pauli operators and $L$ is the system size. 
This model involves nearest-neighbor interactions and no site inhomogeneities. We consider (anti-)periodic boundary conditions for even (odd) $L$ hereafter for more consistent performance comparison unless specified otherwise (see Supplementary Note 2). Starting from the ground state of  $ H(0) = h_{zi} \sum_{j=1}^L \sigma^z_j$, we focus on the task of preparing ground state of the transverse-field Ising model (TFIM) $H(\tau) = \sum_{j=1}^L (h_{xf} \sigma^x_j + J_{f}\sigma_{j}^{z}\sigma_{j+1}^{z})$ within a finite time $\tau$. More general cases are in Supplementary Note 5.
We assume $h_{zi} = J_f = 1$, $h_{xf}>0$, and $J_f \tau=1$ hereafter unless specified otherwise. 
The choice $J_f \tau=1$ corresponds to an intermediate-time, finite-speed regime that is relevant for near-term hardware.
The TFIM features a phase transition from antiferromagnetic ordering $h_{xf}<1$ to a disordered phase $h_{xf}>1$.
The Hamiltonian can be decomposed into a series of independent Landau-Zener Hamiltonians after Jordan-Wigner transformation. While in the fermion operator form the counterdiabatic driving is known, in the spin picture the exact CD Hamiltonian $H_{\mathrm{CD}}$ is highly non-local involving multi-body terms with long-range interaction~\cite{delCampoPRL2013,delCampoPRL2012,delCampo2015,WurtzPRB2020} (see also Supplementary Note 2).
To eliminate the need for these inaccessible ancillary controls and avoid using explicit spectral information to construct the shortcut to adiabaticity, we adopt the variational approach described above to find the time-varying control field using a single-body ansatz.

Specifically, we consider the smooth function $\lambda(t) = \sin^{2}(\frac{\pi}{2}\sin^{2}(\frac{\pi t}{2\tau}))$  as a sweep function~\cite{Kolodrubetz2017} from $t=0$ to $t=\tau$ with $\lambda(0) = 0$ and $\lambda(\tau) = 1$.
Its first and second derivatives vanish at the beginning and the end of the protocol.
The adiabatic ground state preparation process is given by 
\begin{equation}\label{eq:adiabatic_Ht}
    H_0 (t) = \sum_{j=1}^L \left[ (1-\lambda(t)) \sigma^z_j + \lambda(t) \left( h_{xf} \sigma^x_j +  J_f \sigma_{j}^{z}\sigma_{j+1}^{z} \right)  \right].
\end{equation}
As the matrix of Hamiltonian $H(t)$
is real-valued,  $A_\lambda$ needs to be imaginary in this case and we can choose the approximate gauge potential
that contains terms as products of an odd number of $\sigma^{y}$ with $\sigma^{x(z)}$.
The simplest example would be local $\sigma^y$ terms that correspond to the first-order approximate gauge potential, that is, $A_{\lambda}^{(1)}= \sum_{j=1}^L \alpha\sigma_{j}^{y}$.  The parameter $\alpha$
can be found variationally by minimizing the action $S(A_{\lambda}^{(1)})$ (see Methods and Supplementary Note 4 for details). The final time-dependent Hamiltonian is thus 
\begin{equation}\label{eq:LCD_Ht}
    H_{\mathrm{LCD}}(t) = H_{0}(t) + \lambda_f \sum_{j=1}^L \dot{\lambda} \alpha(t) \sigma^y_j,
\end{equation}
where $\lambda_f$ is a free parameter to tune.
In the following, we will show that this simple single-body operator could efficiently suppress diabatic transitions in specific model parameter regimes.

\subsection{Optimization of LCD control}
\begin{figure*}
  \centering
  \includegraphics[scale=0.53]{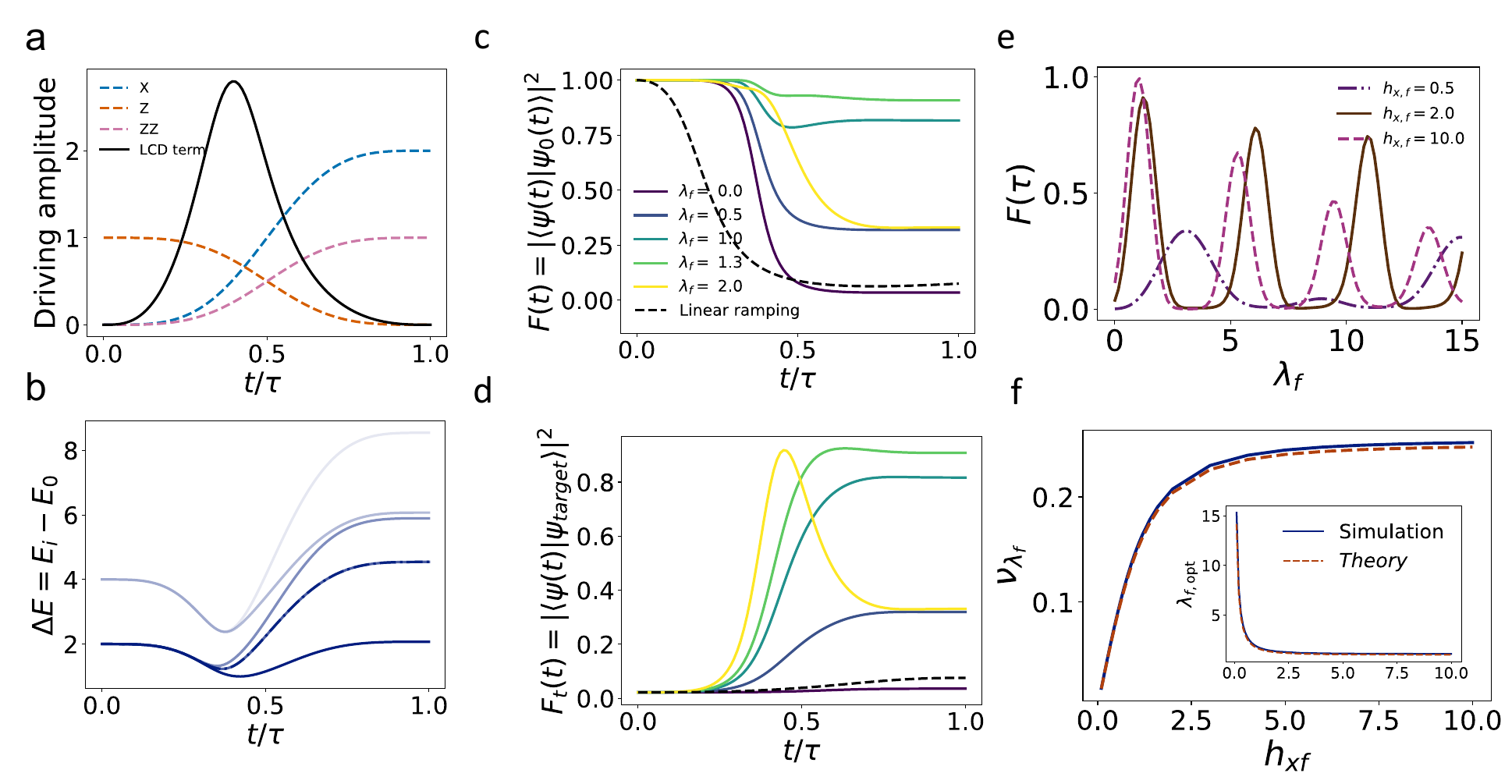} 
  \caption{ \textbf{Time evolution and performance optimization of local counterdiabatic driving.}  \textbf{a.} Time evolution of the driving amplitudes for $h_{xf} = h_x(\tau) = 2$. The LCD term corresponds to local $Y$ operators with $\lambda_f = 1.3$. \textbf{b.} Instantaneous energy spectrum of $H_0(t)$ for $h_{xf} = 2$. Energy difference between the first six excited states and the ground state is shown. The second and third excited states are degenerate here. \textbf{c-d.} Fidelity of the system state $\ket{\psi(t)}$ with respect to the instantaneous ground state $\ket{\psi_0(t)}$ of $H_0(t)$ (c) and target state $\ket{\psi_{\mathrm{target}}}$ to be prepared (d) for $h_{xf} = 2$. The $\lambda_f=0$ curve is the  baseline protocol using the sweep function specified in Eq.~\ref{eq:adiabatic_Ht}, while the linear ramping curve corresponds to a linear adiabatic control
  $H_{\mathrm{linear}}(t)= (1-t/\tau)\sum_{j=1}^L \sigma^z_j + t/\tau  \sum_{j=1}^L (h_{xf}\sigma^x_j + \sigma_{j}^{z}\sigma_{j+1}^{z})$.
  \textbf{e.} Final fidelity with respect to the target state  with scans of the LCD protocol parameter $\lambda_f$ for different model parameters $h_{xf}$. \textbf{f.} Fidelity oscillation frequency $\nu_{\lambda_f}$ as a function of $h_{xf}$, found by both numerical simulations and theoretical analysis. The inset figure shows the optimal $\lambda_f$ for different $h_{xf}$ values. The system size is $L=4$ for plots in this figure and the results hold qualitatively for larger system sizes. \label{fig:main_fig1}} 
\end{figure*}

We begin by presenting the dynamics of LCD evolution.
 The time-dependent LCD control field is shown in Fig.~\ref{fig:main_fig1}a at $h_{xf}=2$. The instantaneous spectrum of $H_0(t)$ shows a small energy gap between the ground and first excited  states in the middle of evolution (Fig.~\ref{fig:main_fig1}b), where the LCD driving amplitude is largest.
 This gap indeed increases the transition rate from the instantaneous ground state to excited states, thereby reducing the fidelity of the  baseline driving protocol (which corresponds to $\lambda_f=0$).

 To benchmark the performance of the LCD protocol, we first present the instantaneous fidelity of the state from the evolution of $H_{\mathrm{LCD}}(t)$ with respect to the ground state of $H_0(t)$ in Fig.~\ref{fig:main_fig1}c for various driving amplitudes $\lambda_f$, \textit{i.e.}, $F(t) = \abs{\bra{\psi(t)}\ket{\psi_{0}(t)}}^2$: as expected, the evolution following $H_0(t)$ in Eq.~\ref{eq:adiabatic_Ht} (solid blue curve) or following a sweep function linear in time (dashed black curve) both show low fidelity at $t=\tau$ due to the limited evolution time  (see Supplementary Note 5C for benchmark results of varying $J_f \tau$). 
Throughout, we refer to this reference protocol as the
\emph{baseline sweep}: the standard interpolation under $H_0(t)$ executed in the finite total time $\tau$ with no CD term, for which only in the limit $\tau\to\infty$ would its fidelity approach unity.
 The LCD evolutions, on the other hand, have only minor fidelity reductions induced by diabatic transitions. This indicates that the LCD driving, although it only involves the first-order approximate adiabatic gauge potential, can efficiently suppress the diabatic transitions in the regime $h_{xf}>1$ studied here. 
 Physically, at this intermediate evolution time, the dominant diabatic response is still well captured by the leading operator in the variational AGP ansatz, whereas at much shorter times higher-order operator structures become important and any fixed low-locality ansatz necessarily loses effectiveness (see Supplementary Note 5C).

Interestingly, our findings indicate that, for the specified model and with certain parameters, the LCD schedule achieves high fidelity relative to the target state mid-evolution, but exhibits a low fidelity at the end of evolution. An example is shown in Fig.~\ref{fig:main_fig1}d, where the fidelity $F_t(t) = \abs{\bra{\psi(t)}\ket{\psi_{\mathrm{target}}}}^2$ reaches its maximum at $t\approx 0.44 \tau$ when $\lambda_f= 2.0$. Indeed, as $\lambda_f$ increases, the LCD term dominates the evolution and the time to reach maximum fidelity decreases. 
This indicates that, for some $\lambda_f$, the overlap with the target ground state peaks at an intermediate time $t^\star<\tau$. This observation suggests that an early-termination (truncated) sweep or a schedule explicitly designed to end at $t^\star$ might further reduce the required runtime. A systematic time-optimal analysis that connects shortcut protocols and minimal-time bounds~\cite{Puebla2020PRR} is beyond the scope of this work.

The $\lambda_f$-dependent performance motivates us to further investigate how the final fidelity varies with this parameter, aiming to identify the optimal setting. In Fig.~\ref{fig:main_fig1}e, the fidelity at $t=\tau$ shows an oscillatory behavior as a function of $\lambda_f$, and the oscillation frequency changes with $h_{xf}$.
This approximate periodicity in the driving amplitude can be understood in a rotating frame where the added $\sum_j \sigma^y_j$ LCD terms are canceled out (see Methods and Supplementary Note 3). Specifically, we find that the oscillation frequency is given by
\begin{equation}
    \nu_{\lambda_f} = \frac{1}{\pi} \int_{0}^{\tau} d t' \, \dot{\lambda} (t') \alpha (t'),
\label{eq:nu_lambda_f}
\end{equation}
which can be directly computed from the analytical formulas of $\lambda(t)$ and $\alpha(t)$. 
Furthermore, we theoretically identify the optimal $\lambda_f$ as $\lambda_{f, \mathrm{opt}} = \frac{1}{4 \nu_{\lambda_f}}$ as shown in Fig.~\ref{fig:main_fig1}f. We note that the least action principle with the LCD ansatz in Eq.~(\ref{eq:LCD_Ht}) would analytically yield $\lambda_f = 1$, which minimizes the total transition rate from \textit{any} eigenstate to all other states~\cite{Kolodrubetz2017}. 
In contrast, our objective is to prepare only the ground state rather than minimizing the total transition rates, that is, we would have $\lambda_{f, \mathrm{opt}} \neq 1$ in general (see Methods for details).

We note a significant increase in optimal $\lambda_f$ as $h_{xf}$ decreases. This can be intuitively explained by the following. The minimal energy gap between ground state and first excited state during the evolution gets smaller and smaller (note that the ground states become degenerate in the thermodynamic limit at $h_{xf}<1$). From Eq.~\ref{eq:exact_AGP} and Eq.~\ref{eq:exact_AGP_approximate}, we see that to make the introduced LCD-term a better approximation of the exact AGP, a larger driving amplitude $\alpha$ is necessary to match the coefficients $\frac{1}{\epsilon_m-\epsilon_n}$ in exact AGP and $\alpha_1 (\epsilon_m - \epsilon_n)$ in approximate gauge potential. In essence, diminishing energy gaps generally necessitate larger LCD driving amplitudes.

\subsection{LCDLU}
\begin{figure*}
  \centering
  \includegraphics[scale=0.55]{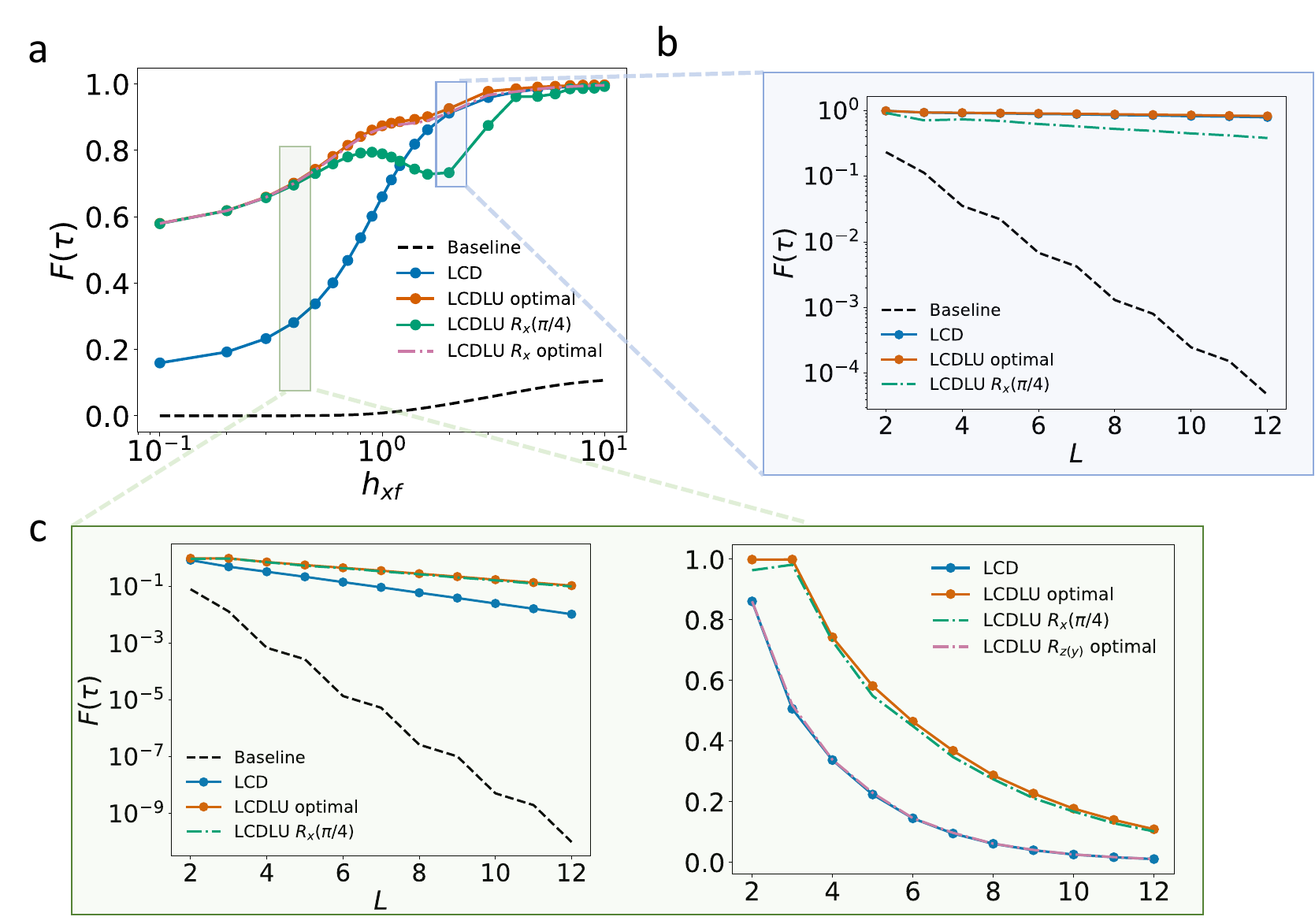}
  \caption{\textbf{Performance of LCD and its improvement.} \textbf{a.} Fidelity of  baseline driving, LCD driving and proposed LCDLU for $L=4$. The LCD driving amplitude parameter $\lambda_f$ is optimized for each $h_{xf}$. The optimal LCDLU curve signifies the identification of the optimal local unitary through numerical exploration across all potential single-body unitary evolutions and serve as an upper bound. Conversely, the $R_x$ optimal curve represents the highest fidelity achieved when local unitary operations are confined to global $X$-rotations. The green line indicates the LCD driving followed by a global $X$-rotation with fixed angle $\pi/4$. \textbf{b.} Fidelity as a function of the system size for $h_{xf}=2$.  \textbf{c.} Fidelity as a function of system size for $h_{xf}=0.5$. The right figure shows the same comparison in linear scale. The $R_{z(y)}$ optimal curve represents the best fidelity achieved when only allowing $Z(Y)$-rotations for LU. We fit the curves with a single exponential function $F(\tau)$=$2^{-cL+a}$ for different protocols to extract the scaling exponent $c$.
   \label{fig:main_fig2}}
\end{figure*}

We next benchmark the performance of LCD driving protocol across different model parameter regimes. We therefore vary $h_{xf}$ and find the state preparation fidelity at end of evolution $t=\tau$ with optimal LCD driving amplitude $\lambda_f$. Fig.~\ref{fig:main_fig2}a shows that LCD can consistently surpass  the baseline driving, particularly when $h_{xf}$ is large.
Indeed, in the limit of large $h_{xf}$, the desired target state closely approximates the ground state of $\sum_j \sigma^x_j$, indicating that the target Hamiltonian closely resembles a trivial single-body model.
The optimal operation is then simply close to $\sum_j\sigma^y_j$ that rotates the ground state of $\sum_j \sigma^z_j$ to $\sum_j \sigma^x_j$. The LCD driving protocol approaches optimality, thereby yielding high fidelity that is close to unity and is much better than standard adiabatic evolution. 

However, the fidelity drops down for small $h_{xf}$, particularly in the regime where the target state is the ground state of TFIM in the ordered phase ($h_{xf}<1$). This highlights the inadequacy of local single-body drivings. In this scenario, due to the vanishing energy gap between the ground state and the first excited state, more complicated driving mechanisms are needed to suppress diabatic transitions.
Indeed, the observed phenomenon is connected to the dominance of two-body interactions in the Hamiltonian $H_0(t)$ when $h_{xf} < 1$. 
This condition necessitates non-local corrections, which are present in higher-order approximate adiabatic gauge potentials, to cancel diabatic transitions efficiently. Specifically, 
we find that the $l=2$ order approximate gauge potential following Eq.~\ref{eq:approx_AGP} contains the following two-body terms (see Supplementary Note 4):
\begin{equation}\label{eq:2nd_order_LCDterms}
  \sum_j (\sigma^y_j \sigma^z_{j+1} + \sigma^z_j \sigma^y_{j+1}), \quad \sum_j (\sigma^y_j \sigma^x_{j+1} + \sigma^x_j \sigma^y_{j+1}).
\end{equation}
Directly including these terms in the original counterdiabatic driving protocol would be more costly than just having single-body driving. 
We therefore consider designing a  counterdiabatic protocol by leveraging the state prepared by LCD protocol which already has a considerably better performance than the standard adiabatic approach.

We aim to reach the target state with high fidelity by appending a simple local unitary (LU) at the end of the LCD sweep, namely a single depth-1 layer of identical single-qubit rotations applied to all spins, $\mathrm{LU} = R_{\hat n}(\theta)^{\otimes L}$ (and in particular $R_x(\theta)^{\otimes L}$ in the TFIM case studied here).
In Fig.~\ref{fig:main_fig2}a, 
we demonstrate that the performance of LCDLU surpasses that of LCD, particularly when $h_{xf}$ is small. In this regime, the optimal rotation axis and angle for LU closely align with an $X$-rotation of $\pi/4$. This is substantiated by comparing the performance using the optimal local unitary, determined through numerical investigation of all potential single-body rotations, against those obtained using fixed $\mathrm{LU} = \exp(-i \frac{1}{2} \frac{\pi}{4} \sum_j \sigma^x_j)$. 
When $h_{xf}$ is large, given that LCD driving already attains high fidelity in preparing target states close to the product states $\ket{-X}^{\otimes L}$, the inclusion of a second-step rotation offers minimal benefit and adding fixed-angle LU ($X$-rotation with angle $\pi/4$) would not change the state. In the intermediate $h_{xf}$ range, the fixed-angle LU would lead to worse performance. This is because the LCD-prepared state is already close to the target state and the fixed-angle LU would reduce the target state population in it, as the target ground state is no longer the eigenstate of fixed-angle LU.
In Fig.~\ref{fig:main_fig2}a we show both an optimal LCDLU curve obtained by numerical optimization over single-qubit rotations (used only as a performance upper bound), and a simple experimentally motivated choice, $\mathrm{LU}=R_x(\pi/4)^{\otimes L}$, which we use in all following experiments and which does not require any fidelity-based optimization or diagonalization.

 We next discuss the physical intuitions behind the improved performance given by the fixed-angle LU in small $h_{xf}$ regime.
Firstly, a heuristic picture is that the combination of first-step LCD followed by an $X$-rotation LU generates an effective evolution that contains $[H_{\mathrm{LCD}},\sum_j\sigma^x_j]$ where the $ \sum_j (\sigma^y_j \sigma^z_{j+1} + \sigma^z_j \sigma^y_{j+1})$ term in Eq.~\ref{eq:2nd_order_LCDterms} shows up. 
This is indeed the  dominant two-body term: see Supplementary Note 4B for details, including results showing that LCDLU can approach two-body LCD in the intermediate regime while avoiding additional two-body control. 
Furthermore, appending $U_{LU}$ leaves the LCD evolution and initial state unchanged and is exactly equivalent to scoring the same output $|\psi(\tau)\rangle$ against the ground state of a conjugated target, $|\langle\psi_{\rm target}|U_{LU}|\psi(\tau)\rangle|^2=|\langle{\rm GS}(H^{(\theta)}_f)|\psi(\tau)\rangle|^2$ with $H^{(\theta)}_f=U^\dagger_{LU}H_f U_{LU}$ and $H_f=H(\tau)$ is the target Hamiltonian. For a global $X$-rotation the interaction is redistributed, $U^\dagger_{LU}\sigma^z_j\sigma^z_{j+1}U_{LU}=\cos^2\theta\,\sigma^z_j\sigma^z_{j+1}+\sin\theta\cos\theta(\sigma^y_j\sigma^z_{j+1}+\sigma^z_j\sigma^y_{j+1})+\sin^2\theta\,\sigma^y_j\sigma^y_{j+1}$, generating exactly the $\sum_j(\sigma^y_j\sigma^z_{j+1}+\sigma^z_j\sigma^y_{j+1})$ channel that dominates the second-order gauge potential but is missing from the single-body ansatz. The deformation thus trades the part of the $\sigma^z\sigma^z$ interaction that single-body LCD cannot follow for the operator content it does generate, giving a target whose ground state the LCD output reaches with higher fidelity. The axis is fixed, not fitted: $|\psi(\tau)\rangle$ and the target ground state share the parity $\prod_j(-\sigma^x_j)$, and the optimal $Y$- or $Z$-rotation cannot beat bare LCD, leaving only the angle, with near-optimal value $\theta\approx\pi/4$ from a single calibration. Supplementary Note 4B gives a complementary dynamical derivation and a direct comparison showing fixed-angle LCDLU approaches two-body LCD using only a terminal single-body rotation.

To further benchmark the performance, we evaluate the scalability of LCD and LCDLU (for both true optimal and fixed-angle LU) protocols relative to system size. Fig.~\ref{fig:main_fig2}b describes the scaling of fidelity at $h_{xf}=2$, at which the LCD and LCDLU show similar results with an exponential fitting $F(\tau)=2^{-cL+a}$ with $c\approx 0.02$ and $0.03$, respectively, while the standard adiabatic approach gives $c_a\approx 1.18$. At small $h_{xf}$ values,  the scaling exponent for adiabatic evolution increases rapidly, approaching a value of $c_a\approx 2.65$ at $h_{xf}=0.5$. In this case, the LCDLU (both true optimal and fixed-angle LU) yields $c_{lu}\approx 0.31$, whereas for LCD we get $c\approx 0.66$ (Fig.~\ref{fig:main_fig2}c). Indeed, at small $h_{xf}$, the energy gap between the ground state and the first excited state is exponentially small in the system size $L$ and scales as $\Delta E \sim (h_{xf}/J_f)^L$. Although the LCD-based protocols still suffer from the  exponential decay in fidelity due to vanishing energy gap as system size increases, the proposed LCDLU demonstrates a better scaling behavior than both LCD and the adiabatic approach. 
The fitted exponents here should be interpreted as effective values at accessible system sizes, not proven asymptotic scalings.
Theoretical proof that LCDLU can provide a scaling advantage beyond the heuristic evidence here would be an interesting topic for future work.

The proposed two-stage protocol comprises an evolution with single-body-based LCD driving and a following local unitary operation, without necessitating additional non-local multi-body terms, thereby making it more accessible for experimental implementation. 
We emphasize that identifying a near-optimal LU \emph{form} (axis) and a good \emph{approximate} angle can be guided by symmetry and the structure of the next-order approximate AGP, without diagonalization; in practice, any remaining fine-tuning reduces to scanning a small number of experimentally accessible parameters (e.g., a single angle for $R_x(\theta)^{\otimes L}$).
As one of the main results of this work, this observation, coupled with the optimal LCD free parameter $\lambda_f$ identified through our theoretical analysis, enables the entire LCDLU process to enhance fidelity without requiring computationally intensive calculations on system dynamics and numerical optimization of driving parameters.

\subsection{Experimental demonstrations}
To demonstrate the efficiency of the proposed method, we perform  digitized implementation of the LCD protocol followed by local single-body driving (LU) in Quantinuum H1-1 trapped-ion system up to system size $L=14$. Recent advances in trapped-ion platforms based on the
QCCD architecture~\cite{Pino2021,Moses2023RaceTrack}  have led to demonstration of deep quantum circuits with increasing number of qubits and with high-fidelity operations.  
While the TFIM studied in this work involves only nearest-neighbor interactions and interaction between boundary qubits, 
the system allows high-fidelity two-qubit gates (with typical gate error $2\times 10^{-3}$) between arbitrary pairs of qubits with low crosstalk through transporting ions into physically separate gate zones, thus allowing us to implement the periodic boundary conditions.

\begin{figure*}
  \centering
  \includegraphics[scale=0.53]{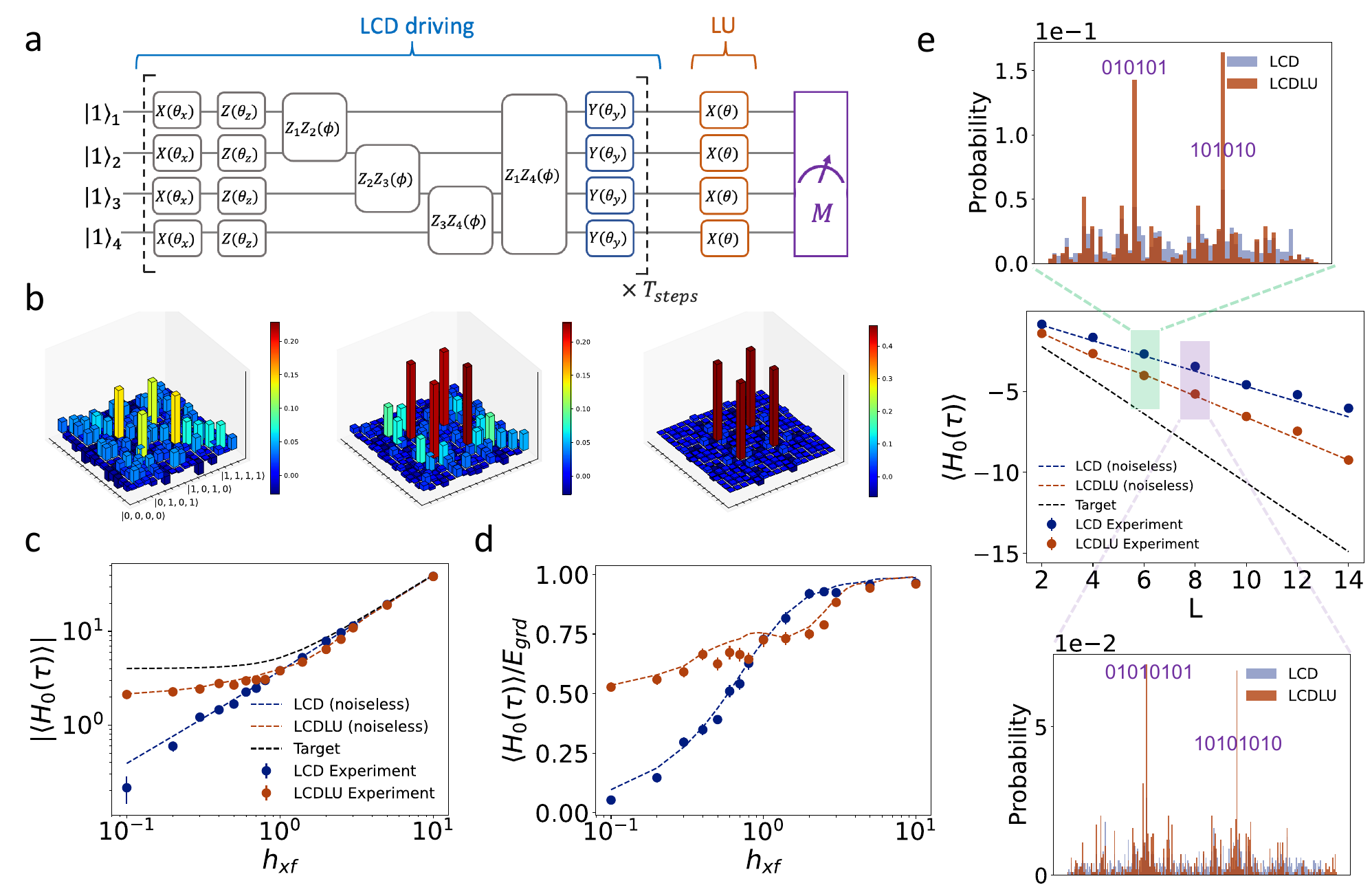}
  \caption{\textbf{Experimental demonstration of LCD driving protocols.} \textbf{a.} Circuit diagram for digitized implementation of LCD and LCDLU using $L=4$ as an example. The number of Trotter steps is specified by $T_{steps}$, and the measurements can be conducted in different bases to extract distinct information about the final state. The evolution parameter $\theta_{x(y,z)}$, $\phi$ is determined by the LCD protocol in Eq.~\ref{eq:LCD_Ht}.  \textbf{b.} Real part of the constructed density matrices for LCD (left) and LCDLU (middle) from four-qubit state tomography experiments and calculated target state density matrix (right), with $h_{xf}=0.5$ and $\lambda_f = 3.095$.  \textbf{c.} Expected energy $|\langle H_0 \rangle|$ as a function of $h_{xf}$ with $L=4$. The dashed lines shows the noiseless simulation results as well as energy for the target ground state ($E_{grd}$). \textbf{d.} Measured energy ratio with respect to energy of the target state. \textbf{e.} Performance of LCD and LCDLU for different system sizes up to $L=14$. The numerical results (dashed lines) for the plot in the middle are from solving the system dynamics via exact diagonalization. The upper (lower) plot shows the probability distribution of Z-basis measurement results for $L=6$ ($L=8$).
  The Trotter step is chosen to be $T_{steps} = 20$ for all the results shown here. The repetition number of circuit run is 1000
  for expected energy measurement and 400 for each circuit in state tomography. Error bars represent one standard error, propagated from the statistical uncertainty in the measured Pauli expectation values. Additional experimental results are shown in Supplementary Note 7. \label{fig:main_fig3}}
\end{figure*}

We implement the digitized LCD protocol described in Eq.~\ref{eq:LCD_Ht} as well as the fixed-angle LCDLU protocol
through Trotterization~\cite{Barends2016,Hegade_2022,PhysRevA.104.L050403}, as shown in the circuit in Fig.~\ref{fig:main_fig3}a. The Trotter step is chosen to approximate the continuous evolution while reducing the circuit depth to prevent overhead in gate counts and reduce gate errors (see Supplementary Note 7 for details). In our experiments, the LU rotation is a fixed $R_x$ rotation with an angle of  $\pi/4$. In order to extract the final fidelity at the end of protocol $t=\tau$, we first perform quantum state tomography to reconstruct the full density matrix. In Fig.~\ref{fig:main_fig3}b, we present the experimental results for the constructed density matrix for a four-qubit system with $h_{xf} = 0.5$ for both LCD and LCDLU protocols. 
In this case, the ground state of the system is close to $\frac{1}{\sqrt{2}}(\ket{1010}+\ket{0101})$, indicative of the antiferromagnetic ordered phase. It is shown that the fixed-angle $\mathrm{LU}=  \exp(-i \frac{1}{2} \frac{\pi}{4} \sum_j \sigma^x_j)$ can effectively increase the corresponding entries in the density matrix, leading to a higher fidelity. 

To further benchmark the performance of both protocols in preparing ground state of different TFIM model parameters $h_{xf}$,
instead of performing costly state tomography, we extract the expected energy of the model $\langle H_0 \rangle = \sum_j (J_f \langle\sigma^z_j \sigma^z_{j+1} \rangle + h_{xf}\langle \sigma^x_j \rangle)$ by performing measurements in $Z$ and $X$ basis. 
We emphasize that although the average energy may serve as an approximate metric for state fidelity at $L=4$, this correlation does not hold for larger system sizes where even minor variations in the system could result in states that are orthogonal to each other despite having similar energies~\cite{PhysRevLett.124.110601}. 
Nonetheless, in Fig.~\ref{fig:main_fig3}c-d we present the experimental results of LCD and LCDLU for a system containing four qubits. Both protocols exhibit small deviations from noiseless simulations. As anticipated, we observe enhanced performance for fixed-angle LCDLU compared against LCD in the regime where $h_{xf}\lesssim1$, as evidenced by the improved energy ratio relative to the target ground state energy. When $h_{xf}\gtrsim5$, both protocols yield an energy that closely approximates the true ground state energy.

To explore the scaling of errors, we vary the system size from 2 qubits to 14 qubits at fixed $h_{xf}=0.5$  (Fig.~\ref{fig:main_fig3}e). Thanks to the high fidelity of two-qubit gate in the trapped-ion system, the experimental results match well with the noiseless simulation. The added LU can consistently lower the energy of the system. We further present the distribution of measurement results in the $Z$-basis for $L=6$ and $L=8$, highlighting that LCDLU can consistently produce bitstrings exhibiting antiferromagnetic ordering with a considerably higher probability than LCD and random bitstring sampling.

\section*{Discussions}

We demonstrate that the optimal
parameters for LCD evolution, as well as the rotation axis and an approximate rotation angle of LU, can be identified without relying on exact diagonalization or extensive numerical optimization. 
This makes the protocols more practically relevant and experimentally usable without the necessity of optimizing time-dependent local control parameters~\cite{COLD_PRXQuantum.4.010312, morawetz2024efficient}.
Although our analysis focuses primarily on the ground state preparation of TFIM with nearest-neighbor interaction, 
the approach of adding simple model-inspired local control to effectuate non-local counterdiabatic corrections during the evolution is broadly applicable to ground state preparation for a variety of models.
More concretely, the LU construction follows a systematic procedure applicable to generic many-body Hamiltonians. Given the single-body LCD ansatz, one identifies the leading operator structures present in the next-order nested-commutator expansion of the approximate gauge potential but absent from the single-body ansatz. The LU generator is then chosen so that its commutator with the LCD Hamiltonian overlaps with those omitted operators.
The rotation angle reduces to a single-parameter calibration that can be performed experimentally without diagonalization.
To further validate the protocol, we additionally study a nonintegrable Ising chain with longitudinal fields $h_z^f \neq 0$, as well as a classical Ising model with all-to-all couplings in Supplementary Note 5D and Note 6; these examples show that LCDLU is not TFIM-specific, but that its gain is controlled by how well the dominant omitted AGP operator can be captured by a simple uniform rotation.
We also remark that extending the protocol to richer local controls for more Hamiltonian families is a natural next step, in particular as a handle on
entanglement and correlation spreading during state preparation.
This protocol, requiring only local control, simplifies experimental implementation and improves on the impractical or costly non-local correction terms that are commonly used.
While we show gate-based  digitized implementation in our experiments here, we remark that the protocols shown here can be more naturally applied in analog quantum systems for ground state preparation. An example in Rydberg neutral-atom arrays  using Rydberg-dressed states is presented in the Methods section.

Furthermore, the protocols based on LCD are directly applicable to a wide range of classical and quantum optimization problems~\cite{PhysRevApplied.20.014024,Hegade_2022,cadavid2023portfolio,guan2023singlelayer}. In fact, CD has been introduced into the Quantum Approximate Optimization Algorithm (QAOA)~\cite{hogg2000quantum,farhi2014quantum,Zhou2020PRX} and demonstrated to improve its performance
~\cite{Wurtz2022CDQAOA,PRA2018_fastOsciallation,PhysRevResearch.4.013141,Vizzuso2024NJP}. 
These works incorporate CD into QAOA either by modifying the QAOA parameters~\cite{Wurtz2022CDQAOA,PRA2018_fastOsciallation} or by adding extra terms to the QAOA circuit~\cite{PhysRevResearch.4.013141}.
It might be interesting to study the possibility of improving QAOA performance by modifying diabatic transitions through extra gates such as the LU technique here.

\section*{Conclusions}

In summary, we introduce and experimentally validate a two-stage shortcut protocol for many-body ground-state preparation that uses only simple, experimentally accessible controls. Beyond standard first-order LCD, we provide an analytically motivated prescription for choosing the global LCD amplitude rescaling \(\lambda_f\) that targets ground-state fidelity without requiring instantaneous eigenstates or diagonalization. We then augment the sweep with a single global on-site rotation (LCDLU), which compensates leading missing gauge-potential contributions at essentially no additional two-qubit gate cost. Using digitized implementations on a trapped-ion processor, we benchmark both LCD and LCDLU and observe consistent performance improvements with system sizes up to 14 qubits. This exploration of local control as a quantum shortcut to adiabaticity paves the way for the preparation of the ground state in many-body systems using readily available controls.

\section*{Methods}

\textbf{Control amplitude of $A_\lambda^{(1)}$ from least action principle.}
The time-dependent driving control $\alpha(t)$ in LCD control specified in Eq.~\ref{eq:LCD_Ht} can be found by minimizing the action discussed above based on the least action principle. Specifically, the operator $G(A_\lambda^{(1)})$ is given by
\begin{equation*}
\begin{aligned}
G(A_{\lambda}^{(1)}) =& \partial_{\lambda}H + i \comm{A_{\lambda}^{(1)}}{H} \\
 & = (h_{z}' + 2\alpha h_{x})\sum_j \sigma_{j}^{z} +  (h_{x}' - 2\alpha h_{z}) \sum_j\sigma_{j}^{x}   \\
&+J'\sum_j\sigma_{j}^{z}\sigma_{j+1}^{z} + 2\alpha J \sum_j (\sigma_{j}^{x}\sigma_{j+1}^{z} + \sigma_{j}^{z}\sigma_{j+1}^{x}),
\end{aligned}
\end{equation*}
where the derivative is with respect to $\lambda$.
The goal is then to minimize the action that is given by the Hilbert-Schmidt norm of the above operator
\begin{equation*} 
\begin{aligned}
\frac{1}{L 2^{L}}\tr \left( G \left(A_{\lambda}^{(1)}\right)^{2} \right)
& = ((h_{z}' + 2\alpha h_{x})^{2} \\
& +  (h_{x}' - 2\alpha h_{z})^{2} +(J')^{2} + 4(\alpha J)^{2}),
\end{aligned}
\end{equation*}
which leads to optimal $\alpha$:
\begin{equation} \label{eq:alpha_anaytical}
\alpha = \frac{1}{2} \frac{-h_{z}' h_{x}+ h_{x}' h_{z} }{h_{z}^{2} + h_{x}^{2} + 2J^{2}}.
\end{equation}
The optimal time-dependent driving amplitude $\alpha(t)$ identified is applied in both our simulations and experiments throughout the paper.
Note that evaluating Eq.~\ref{eq:alpha_anaytical} only requires the operator form and time-dependent couplings in $H_0(t)$ (and their $\lambda$-derivatives), but not eigen-decomposition or spectral-gap information of $H_0(t)$.

We note that the least-action principle used to obtain $\alpha(\lambda)$ minimizes the Frobenius-norm action $\mathrm{Tr}\,G^2$ (a spectrum-averaged criterion). One may alternatively define a state-specific objective, e.g. minimizing $\langle \psi_0(\lambda)|G^2|\psi_0(\lambda)\rangle$, which more directly targets transitions out of the instantaneous ground state. While such a choice can further improve performance, it requires access to $\ket{\psi_0(\lambda)}$ along the path and thus generally relies on diagonalization or costly classical simulation for large systems. In contrast, in this work we have kept the analytically accessible least-action $\alpha(\lambda)$ and introduced a single global rescaling $\lambda_f$, whose near-optimal value can be predicted from our rotating-frame analysis (Eq.~\ref{eq:nu_lambda_f}) without spectral information.
We further remark that this variational method can be further improved by using the Lanczos algorithm in the operator basis~\cite{Takahashi_2024arXiv,bhattacharjee2023lanczos,PhysRevX.13.011041}.

\textbf{Optimization of LCD driving}
We consider a time-dependent rotating frame with a uniform $Y$-rotation defined by $U_y(t) = \otimes_i R_y (\theta(t)) = e^{-i \theta_y(t) \sum_j \sigma_j^y/2}$ where $\theta_y(t) = 2 \lambda_f \int_{0}^{t} d t' \, \dot{\lambda} (t') \alpha (t')$\footnote{Note that this moving frame by $Y$-rotation is different from the moving frame defined by the instantaneous exact eigenstates 
(see, \textit{e.g.}, Ref~\citep{Kolodrubetz2017}).}.
In this rotating frame, the $\sigma^y$-terms can be cancelled exactly, resulting in the following effective Hamiltonian
\begin{equation}
\begin{aligned}
     & \widetilde{H}_{\mathrm{LCD}} (t) = U_y (t)^\dagger H_{\mathrm{LCD}} (t) U_y (t) + i \left( \partial_t U_y (t) ^\dagger \right) U_y (t) \\
     = & \left[ h_x (t) \cos\theta_y(t) - h_z(t) \sin \theta_y(t) \right] \sum_j \sigma^x_j \\
     & + \left[ h_z (t) \cos\theta_y(t) + h_x(t) \sin \theta_y(t) \right] \sum_j \sigma^z_j \\
     & + J(t) \sum_j \left( \sigma^z_{j} \cos \theta_y(t) - \sigma^x_{j} \sin \theta_y(t) \right) \\
     & \otimes \left( \sigma^z_{j+1} \cos \theta_y(t) - \sigma^x_{j+1} \sin \theta_y(t) \right)
    .
\end{aligned}
\label{eq:h_lcd_m}
\end{equation}
At the final time $\tau$, since $\theta_y(\tau)$ is proportional to $\lambda_f$, $\widetilde{H}_{\mathrm{LCD}} (\tau)$ and the ground state in this moving frame would be periodic in $\lambda_f$ with the frequency $\nu_{\lambda_f}$ specified in Eq.~\ref{eq:nu_lambda_f}. In the middle of the time range $0 < t < \tau$, the change of $\widetilde{H}_{\mathrm{LCD}}$ and its ground state would not have the same frequency as $\nu_{\lambda_f}$. As a result, the complete dynamics and $F(\tau)$ as its outcome would not follow a perfect periodicity with $\nu_{\lambda_f}$.
In fact, as $\lambda_f$ grows large, the envelope of the oscillation in $F(\tau)$ starts to decay.
However, when $\lambda_f$ is not too large, in numerical study shown in Fig.~\ref{fig:main_fig1}e-f, we have found that the behavior of $F(\tau)$ with changing $\lambda_f$ matches well with the analysis above. We therefore conjecture that the dominating factor for the observed fidelity oscillation is related to the period in
$\theta_y(\tau)$ in contrary to the details of the intermediate values $\theta_y(t)$. We provide more details and explanations in the Supplementary Note 3.

In noiseless simulation for the continuous time (with numerical discretization), we numerically find 
$\argmax_{\lambda_f} F(\tau)$ using the Brent's method~\cite{brent1971algorithm} and compare with the theoretical analysis. In Fig.~\ref{fig:main_fig2}, due to the computational cost in simulation at $L = 12$ for optimizing $F(\tau)$, we use the $\lambda_{f, \mathrm{opt}}$ found at $L = 11$, as $\lambda_{f, \mathrm{opt}}$ exhibits only slight variation with $L$.

\textbf{Optimization of LU axis and angle.}
After LCD protocol that prepares a state $\ket{\psi(\tau)}$, we apply a parameterized local special unitary operator 
$\mathrm{LU}(\{\alpha_i, \beta_i, \theta_i\}) = \otimes_{i} R_z (\alpha_i) R_x(\theta_i) R_z (\beta_i)$ which is of the most general form with the overall phase omitted~\cite{PhysRevA.52.3457}.
We take this general form with $3 L$ parameters when finding for the optimal LCDLU protocol.
More restrictively, when we assume the translational symmetry in $\mathrm{LU}$, parameters at all the sites $i$ are taken the same, and thus the $\mathrm{LU}$ becomes an uniform rotation $\mathrm{LU}(\alpha, \beta, \theta) = (R_z (\alpha) R_x(\theta) R_z (\beta))^{\otimes L}$.
Further, when the uniform rotation is fixed to the $X$-rotation, $\mathrm{LU} (\theta) = R_x (\theta)^{\otimes{L}}$.
Using the fidelity between the LCDLU state and the ground state as an optimization objective, with the general form of $\mathrm{LU}$, we use the COBYLA~\cite{Powell1994} optimizer to numerically find the optimal LU. In the single-parameter optimization case, we use the Brent's method in optimization to find $\theta$.
Because exact state-vector benchmarking grows exponentially with $L$, the present numerical study emphasizes system sizes for which apples-to-apples fidelity comparisons remain reliable, and complements this with broader model classes and hardware experiments up to $L=14$.

The fidelity-based optimization of $U_{\mathrm{LU}}$ described above is used to benchmark the best attainable LCDLU performance. The practical protocol advocated in the main text uses a AGP- and symmetry-guided choice of LU axis together with either a fixed angle or a one-parameter calibration based on a simple observable such as the final energy.

\textbf{Feasibility on Rydberg neutral-atom arrays.}
As an implementability remark, here we show the feasibility of implementing the LCD-based protocols on Rydberg neutral-atom arrays; the related counterdiabatic protocols tailored to neutral-atom hardware have been proposed, e.g. in~\cite{2405.14829}.
Using Rydberg-dressed states ~\cite{Balewski2014,PhysRevLett.112.103601,Eckner2023Rydberg,PhysRevLett.131.063401}, an effective Ising-form Hamiltonian can be engineered,
\[
H_{\mathrm{Ryd}}(t)=\Omega(t)\sum_j \sigma^x_j + \sum_j \delta_j(t)\sigma^z_j + \sum_{i<j} V_{ij}(t)\sigma^z_i\sigma^z_j,
\]
where $\Omega(t)$ and $\delta_j(t)$ are controlled by laser amplitude and detuning, and dressing allows an effective interaction scale that can be tuned in time, typically as a global prefactor $V_{ij}(t)=v(t)\,V^{(0)}_{ij}(r_{ij})$.
Accordingly, the single-body LCD term (a global $\sum_j \sigma^y_j$ control) can be realized by choosing the drive phase, and the scalar rescaling $\lambda_f$ corresponds to an amplitude rescaling.
The LU step in LCDLU is a single global on-site rotation that can be implemented as a short pulse.

\section*{Data availability}
The data that support the findings of this study are available from the corresponding authors upon
request at \texttt{quantum.computing@jpmchase.com}.

\section*{Code availability}
The code used to generate the simulation results and to run the experimental circuits in this study is available from the corresponding authors upon request at
\texttt{quantum.computing@jpmchase.com}.

\section*{Author contributions}
C.L. conceived the project. C.L. and J.S. performed the numerical simulations and analyzed the results with input from R.S. C.L. carried out the experiments on the H1-1 hardware with support from R.S., J.S. and M.P. All authors discussed the results and contributed to writing the manuscript.

\section*{Funding Declaration}
The authors have no funding to declare.

\section*{Competing interests}
The authors declare no competing interests.

\acknowledgements
The authors thank Yue Sun, Dylan Herman, Shouvanik Chakrabarti, Zichang He, and other colleagues at Global Technology Applied Research of JPMorgan Chase for helpful feedback. C.L. would like to thank Hengyun Zhou and Guoqing Wang for insightful discussions.

\bibliography{draft_arxiv_update} 

\section*{}
\subsection*{Disclaimer}

This paper was prepared for informational purposes by the Global Technology Applied Research center of JPMorgan Chase \&  Co.
This paper is not a product of the Research Department of JPMorgan Chase \& Co. or its affiliates. Neither JPMorgan Chase \& Co. nor any of its affiliates makes any explicit or implied representation or warranty and none of them accept any liability in connection with this paper, including, without limitation, with respect to the completeness, accuracy, or reliability of the information contained herein and the potential legal, compliance, tax, or accounting effects thereof. This document is not intended as investment research or investment advice, or as a recommendation, offer, or solicitation for the purchase or sale of any security, financial instrument, financial product or service, or to be used in any way for evaluating the merits of participating in any transaction.

\clearpage
\onecolumngrid
\clearpage

% Supporting material


\section*{Supplementary material}

# Supplementary material of "Quantum counterdiabatic driving with local control"

Changhao Li,[1, *] Jiayu Shen,[1, *] Ruslan Shaydulin,[1] and Marco Pistoia[1]
[1] *Global Technology Applied Research, JPMorgan Chase, New York, NY 10017, USA*

## CONTENTS

---

* These authors contributed equally to this work. All correspondence should be addressed to changhao.li@jpmchase.com.

# I. BACKGROUND ON QUANTUM COUNTERDIABATIC DRIVING

In this section, we give a more detailed introduction on quantum counterdiabatic driving and its relation to quantum geometry of eigenstates of the Hamiltonian.

## A. Approximate adiabatic gauge potential from least action principle

We start by presenting the details of finding optimal approximate adiabatic gauge potential from the least action principle [1–3].

Consider a non-singular Hamiltonian $H(\lambda)$ parameterized by continuous parameter $\lambda$ (here $\lambda$ could also be a vector containing parameters $(\alpha,\beta,\gamma, ...)$) with eigenstates denoted by $|n(\lambda)\rangle$. If the varying of parameter $\lambda$ is infinitely slow (adiabatic limit), the states $|n(\lambda)\rangle$ are connected by a particular adiabatic unitary transformation and the associated gauge potential $A_\lambda$ is known as adiabatic gauge potential (AGP). Therefore, the AGP is known as the generator of adiabatic deformations between quantum eigenstates. It's diagonal part is indeed the Berry connection. That is, for eigenstate $|n(\lambda)\rangle$ we have

$$A_\lambda^n = i \langle n(\lambda)| \partial_\lambda |n(\lambda)\rangle. \tag{S1}$$

As presented in the manuscript, the AGP connects different eigenstates with energy $\epsilon$ via

$$\langle m| A_\lambda |n\rangle = i \langle m|\partial_\lambda n\rangle = -i\frac{\langle m| \partial_\lambda H |n\rangle}{\epsilon_m - \epsilon_n}. \tag{S2}$$

In the matrix form, the above relations lead to

$$i\partial_\lambda H = [A_\lambda, H] - iM_\lambda, \tag{S3}$$

where the diagonal operator $M_\lambda = -\sum_n \frac{\partial E_n(\lambda)}{\partial \lambda} |n(\lambda)\rangle \langle n(\lambda)|$ is the generalized force operator.

We now can see that the Hermitian operator $G_\lambda(O) = \partial_\lambda H + i[O, H]$ parameterized by operator $O$ is indeed $-M_\lambda$ when the operator $O$ is the exact AGP $A_\lambda$. Therefore, in order to find the optimal term $O$ that approximates the $A_\lambda$, we aim to minimize the Frobenius norm of the operator distance

$$D^2(O) = \text{Tr}\big((G_\lambda + M_\lambda)^2\big) = \text{Tr}\big(G_\lambda^2\big) + \text{Tr}\big(M_\lambda^2\big) + 2\,\text{Tr}(G_\lambda M_\lambda) = \text{Tr}\big(G_\lambda^2\big) - \text{Tr}\big(M_\lambda^2\big). \tag{S4}$$

As $M_\lambda$ does not depend on $O$, we see that the action to be minimized is

$$S = \text{Tr}\big(G_\lambda(O)^2\big), \tag{S5}$$

as mentioned in the main text. We have used this least action principle to find the optimal control for LCD driving.

## B. AGP and quantum geometry

The adiabatic gauge potential, by definition, characterizes the adiabatic deformations between quantum eigenstates. If we consider adiabatic evolution of $H(\lambda)$ with the ground state $|n_0(\lambda)\rangle$, the distance between two nearby states $|n_0(\lambda)\rangle$ and $|n_0(\lambda + d\lambda)\rangle$ with an infinitesimal $d\lambda$ is given by the Frobenius norm of the AGP operator. More specifically, the distance is given by

$$ds^2 = 1 - |\langle n_0(\lambda)|n_0(\lambda + d\lambda)\rangle|^2 = d\lambda_\alpha \chi_{\alpha\beta} d\lambda_\beta + O(|d\lambda^3|), \tag{S6}$$

where $\chi_{\alpha\beta} = \langle \partial_\alpha n_0| (I - |n_0\rangle \langle n_0|) |\partial_\beta n_0\rangle \equiv g_{\alpha\beta} + i\mathcal{F}_{\alpha\beta}/2$ is known as the quantum geometric tensor (QGT) for the ground state $|n_0(\lambda)\rangle$. The real part of QGT $g_{\alpha\beta}$ is the Fubini-Study metric tensor that characterizes the distance between the nearby states and is given by

$$g_{\alpha\beta} = \text{Re}\, \langle n_0| A_{\lambda_\alpha} A_{\lambda_\beta} |n_0\rangle. \tag{S7}$$

The Berry curvature $\mathcal{F}_{\alpha\beta}$ in the imaginary part of QFT is given by the derivatives of the Berry connections (diagonal part of AGP, as in Eq. S1):

$$\mathcal{F}_{\alpha\beta} = i \langle n_0| [A_{\lambda_\alpha}, A_{\lambda_\beta}] |n_0\rangle. \tag{S8}$$

We remark that the distance $ds^2$ above is indeed the probability that the system jumps to excited states when the system parameter is suddenly changed from $\lambda$ to $\lambda + d\lambda$. When the Frobenius norm of the AGP operator is small, indicating negligible distance between nearby states $|n_0(\lambda)\rangle$ and $|n_0(\lambda + d\lambda)\rangle$, the system predominantly remains in its ground state $|n_0\rangle$ following the abrupt change in $\lambda$. Consequently, diabatic transition terms are typically minor, necessitating only minimal amplitude for counterdiabatic driving terms.

## II. TRANSVERSE-FIELD ISING MODEL

In this section, we discuss the background on the transverse-field Ising model (TFIM) with finite $L$ including the choices of boundary conditions and its exact solutions [4, 5]. The major difference between our discussion and the standard derivation from literature is that we consider the general choice of boundary condition including periodic boundary conditions (PBCs) and anti-periodic boundary conditions (ABCs). We give arguments why it is a more natural choice to use the PBCs for even $L$ and ABCs for odd $L$ and consider them as similar classes of problems. The behaviors of the driving protocols in our study of main text would also be more consistent when following these boundary conditions. While many of the results shown below on TFIM can be found in previous literature, we present the discussions on boundary conditions and ground states for the convenience of readers.

In the context of deriving the exact solutions, for convenience of discussion, literature often uses a basis different from our convention of the $x$, $y$, $z$ directions by a rotation in the spin space $\sigma^y$: $\tilde{\sigma}_j^x \equiv e^{-i\frac{\pi}{4}\sigma_j^y}\sigma_j^x e^{i\frac{\pi}{4}\sigma_j^y} = -\sigma_j^z$, $\tilde{\sigma}_j^z \equiv e^{-i\frac{\pi}{4}\sigma_j^y}\sigma_j^z e^{i\frac{\pi}{4}\sigma_j^y} = \sigma_j^x$, $\tilde{\sigma}_j^y \equiv e^{-i\frac{\pi}{4}\sigma_j^y}\sigma_j^y e^{i\frac{\pi}{4}\sigma_j^y} = \sigma_j^y$ where $\tilde{\sigma}$ is for the notation used in this section, and $\sigma$ is for the notation used in our models. The 1D transverse-field Ising model on a ring takes the form

$$H = \sum_{j=1}^{L}\left(J\sigma_j^z\sigma_{j+1}^z + h\sigma_j^x\right) = \sum_{j=1}^{L}\left(J\tilde{\sigma}_j^x\tilde{\sigma}_{j+1}^x - h\tilde{\sigma}_j^z\right). \tag{S9}$$

The spin operators at site 1 and $L+1$ are identified up to a relative phase, and we consider a general boundary condition for the Pauli operators

$$\tilde{\sigma}_{L+1}^\alpha = e^{i\theta_\alpha}\tilde{\sigma}_1^\alpha, \tag{S10}$$

with $\alpha = x, y, z$. The condition for compatible commutation relations of Pauli operators $\left[\tilde{\sigma}_j^\alpha, \tilde{\sigma}_j^\beta\right] = 2i\epsilon_{\alpha\beta\gamma}\tilde{\sigma}_j^\gamma$ for both $j=1$ and $j=L+1$ is

$$\theta_x \in \pi\mathbb{Z}, \quad \theta_y \in \pi\mathbb{Z}, \quad \theta_z \in \pi\mathbb{Z}, \quad \theta_x + \theta_y + \theta_z \in 2\pi\mathbb{Z}. \tag{S11}$$

Thus, the boundary conditions need to satisfy

$$\tilde{\sigma}_{L+1}^\alpha = (-1)^{a_\alpha}\tilde{\sigma}_1^\alpha \tag{S12}$$

with $a_\alpha \in \{0,1\}$ and $a_x + a_y + a_z \equiv 0 \pmod{2}$.

The general procedure to obtain the exact solutions of the 1D TFIM is first mapping the Pauli operators to the hard-core boson operators, and then using the Jordan-Wigner transformation to transform hard-core bosons to spinless fermions. In the end, the fermion Hamitonian is similar to the Bardeen-Cooper-Schrieffer (BCS) theory in 1d, and a Bogoliubov transformation can diagonalize the Hamiltonian which gives the exact solution of the energy eigenstates.

The mapping from Pauli operators and hard-core bosonic operators is

$$\tilde{\sigma}_j^x = b_j^\dagger + b_j, \quad \tilde{\sigma}_j^y = -i\left(b_j^\dagger - b_j\right), \quad \tilde{\sigma}_j^z = 2b_j^\dagger b_j - 1. \tag{S13}$$

and the Jordan-Wigner transformation is

$$b_j = e^{i\pi\sum_{j'=1}^{j-1}n_{j'}}c_j, \quad b_j^\dagger = e^{i\pi\sum_{j'=1}^{j-1}n_{j'}}c_j^\dagger. \tag{S14}$$

Therefore, the mapping between the Pauli operators and spinless fermionic operators is

$$\tilde{\sigma}_j^x = e^{i\pi\sum_{j'=1}^{j-1}n_{j'}}\left(c_j + c_j^\dagger\right), \quad \tilde{\sigma}_j^y = -ie^{i\pi\sum_{j'=1}^{j-1}n_{j'}}\left(c_j^\dagger - c_j\right), \quad \tilde{\sigma}_j^z = 2n_j - 1. \tag{S15}$$

where $n_j = c_j^\dagger c_j = b_j^\dagger b_j$ is the particle number operator at site $j$.

Thus, the boundary condition for hard-core boson operators can be derived from the Pauli operators as

$$b_{L+1} = (-1)^a b_1, \quad b_{L+1}^\dagger = (-1)^a b_1^\dagger \tag{S16}$$

with the requirement $a_z = 0$ and $a_x = a_y$ for compatible mapping rules Eq. S13 at $j=1$ and $j=L+1$. We define $a := a_x = a_y \in \{0,1\}$, as the only parameter in the boundary condition. $a=0$ corresponds to the periodic boundary condition (PBC) of hard-core bosons, and $a=1$ corresponds to the anti-periodic boundary condition (ABC) of hard-core bosons.

Eq. (S14) at $j = L+1$ is

$$b_{L+1} = (-1)^{N_F} c_{L+1}, \quad b_{L+1}^\dagger = (-1)^{N_F} c_{L+1}^\dagger, \tag{S17}$$

where $N_F := \sum_j n_j$ is the total number of fermions (or hard-core bosons, equivalently). Therefore, the boundary condition for the fermion operators is

$$c_{L+1} = (-1)^{N_F+a} c_1, c_{L+1}^\dagger = (-1)^{N_F+a} c_1^\dagger, \tag{S18}$$

where we emphasize that $N_F$ is an operator, and $a \in \{0, 1\}$ is an integer number. The boundary conditions for the Pauli operators in Eq. S12 are not the same as the fermions in Eq. S18. We further note that the fermion parity operator $P = (-1)^{N_F}$ commutes with the TFIM Hamiltonian and thus is a symmetry of the system. The even-$N_F$ and odd-$N_F$ sectors of the Hilbert space do not interact with each other. In the language of Pauli matrices, the parity operator is $P = \exp\left(i\pi \sum_j (\tilde{\sigma}_j^z + 1)/2\right) = \otimes_j(-\tilde{\sigma}_j^z) = \otimes_j(-\sigma_j^x)$. As a remark, the effect of the LU transformation $\mathrm{LU}(\theta) = R_x(\theta)^{\otimes L} = \tilde{R}_z(\theta)^{\otimes L} = \exp\left(-i\frac{\theta}{2}\sum_j \tilde{\sigma}_j^z\right) = \exp(iL\theta/2)\exp(-i\theta N_F)$ on any state is equivalent to unitarily transforming fermion operators as $c_j^\dagger \mapsto e^{i\theta}c_j^\dagger$, $c_j \mapsto e^{-i\theta}c_j$. Note that $\mathrm{LU}(\theta)$ commutes with $P$ and at $\theta = \pi/4$, we have $\mathrm{LU} = R_x(\pi/4)^{\otimes L} = \exp(i\pi L/8)P^{1/4}$.

In the momentum space, the Fourier transform for fermionic operators is

$$c_k = \frac{1}{\sqrt{L}} \sum_{j=1}^L e^{-ikj} c_j, \quad c_k^\dagger = \frac{1}{\sqrt{L}} \sum_{j=1}^L e^{ikj} c_j^\dagger. \tag{S19}$$

The boundary condition of fermion operators in the position space requires $e^{ikL} = (-1)^{N_F+a+1}$ and leads to

$$\begin{aligned} k &\in \frac{2\pi}{L}\mathbb{Z} \quad \text{if } (-1)^{N_F+a+1} = 1, \\ k &\in \frac{2\pi}{L}\left(\mathbb{Z} + \frac{1}{2}\right) \quad \text{if } (-1)^{N_F+a+1} = -1. \end{aligned} \tag{S20}$$

To discuss the different cases of TFIM, we consider the triplet $(L, N_F, a)$, similar to the notation used in Ref. [5], but we have an extra parameter $a$ for the general consideration of boundary conditions. We only consider the parities in $(L, N_F, a)$ instead of their values. We use the notation $E/e$ and $O/o$ for even and odd. $L$ and $a$ define the system. $N_F$ is an operator, but since the fermion parity is a symmetry, it can be used to label the independent sectors. For fixed $(L, N_F, a)$, the possible values of $k$ are given by Eq. S20, and we denote the corresponding set of $k$-values as $\mathcal{K}^{(L,N_F,a)}$.

The last step of exactly solving the TFIM is the Bogoliubov transformation

$$\eta_k = u_k c_k - iv_k c_{-k}^\dagger, \quad \eta_k^\dagger = u_k c_k^\dagger + iv_k c_{-k} \tag{S21}$$

where

$$u_k^2 = \frac{1}{2}\left(1 + \frac{J\cos k + h}{\omega(k)}\right), \quad v_k^2 = \frac{1}{2}\left(1 - \frac{J\cos k + h}{\omega(k)}\right), \quad 2u_k v_k = \frac{J \sin k}{\omega(k)}, \tag{S22}$$

and the quasi-particle energy

$$\omega(k) = \sqrt{(J\cos k + h)^2 + (J\sin k)^2}. \tag{S23}$$

The BCS-like state

$$\left|\mathrm{BCS}^{(L,N_F,a)}\right\rangle := \prod_{\substack{k \in \mathcal{K}^{(L,N_F,a)} \\ 0 < k < \pi}} \left(u_k + iv_k c_k^\dagger c_{-k}^\dagger\right) |\tilde{\downarrow}\rangle^{\otimes L} \tag{S24}$$

is close to the ground state of the TFIM, where $|\tilde{\downarrow}\rangle$ satisfies $\tilde{\sigma}^z|\tilde{\downarrow}\rangle = -|\tilde{\downarrow}\rangle$. The exact ground states depend on $(L, N_F, a)$ and may differ from the BCS-like state. We list the ground states with all possible triplets $(L, N_F, a)$, their energies, and the limit under the $h/J \to 0$ limit when $J > 0$:

$$\left|E_0^{(E,o,0)}\right\rangle = c_\pi^\dagger \left|\mathrm{BCS}^{(E,o,0)}\right\rangle, \quad E_0^{(E,o,0)} = |J+h| - (J-h) - \sum_{k \in \mathcal{K}^{(E,o,0)}} \omega(k) \stackrel{\substack{J>0 \\ h/J \to 0}}{=\!=\!=} -LJ \tag{S25}$$

$$\left|E_0^{(E,e,0)}\right\rangle = \left|\mathrm{BCS}^{(E,e,0)}\right\rangle, \quad E_0^{(E,e,0)} = - \sum_{k\in\mathcal{K}^{(E,e,0)}} \omega(k) \stackrel{\substack{J>0 \\ h/J\to 0}}{=\!=\!=} -LJ \tag{S26}$$

$$\left|E_0^{(O,o,0)}\right\rangle = c_0^\dagger \left|\mathrm{BCS}^{(E,o,0)}\right\rangle, \quad E_0^{(O,o,0)} = |J-h| + (J-h) - \sum_{k\in\mathcal{K}^{(O,o,0)}} \omega(k) \stackrel{\substack{J>0 \\ h/J\to 0}}{=\!=\!=} -(L-2)J \tag{S27}$$

$$\left|E_0^{(O,e,0)}\right\rangle = \eta_{k_*^{(O,e,0)}}^\dagger c_\pi^\dagger \left|\mathrm{BCS}^{(O,e,0)}\right\rangle, \quad E_0^{(O,e,0)} = 2\omega\left(k_*^{(O,e,0)}\right) - \sum_{k\in\mathcal{K}^{(O,e,0)}} \omega(k) \stackrel{\substack{J>0 \\ h/J\to 0}}{=\!=\!=} -(L-2)J \tag{S28}$$

$$\left|E_0^{(E,o,1)}\right\rangle = \eta_{k_*^{(E,o,1)}}^\dagger \left|\mathrm{BCS}^{(E,o,1)}\right\rangle, \quad E_0^{(E,o,1)} = 2\omega\left(k_*^{(E,o,1)}\right) - \sum_{k\in\mathcal{K}^{(E,o,1)}} \omega(k) \stackrel{\substack{J>0 \\ h/J\to 0}}{=\!=\!=} -(L-2)J \tag{S29}$$

$$\left|E_0^{(E,e,1)}\right\rangle = c_0^\dagger c_\pi^\dagger \left|\mathrm{BCS}^{(E,e,1)}\right\rangle, \quad E_0^{(E,e,1)} = -(J+h)+(J-h)+|J+h|+|J-h| - \sum_{k\in\mathcal{K}^{(E,e,1)}} \omega(k) \stackrel{\substack{J>0 \\ h/J\to 0}}{=\!=\!=} -(L-2)J \tag{S30}$$

$$\left|E_0^{(O,o,1)}\right\rangle = c_\pi^\dagger \left|\mathrm{BCS}^{(O,o,1)}\right\rangle, \quad E_0^{(O,o,1)} = |J-h| - (J-h) - \sum_{k\in\mathcal{K}^{(O,o,1)}} \omega(k) \stackrel{\substack{J>0 \\ h/J\to 0}}{=\!=\!=} -LJ \tag{S31}$$

$$\left|E_0^{(O,e,1)}\right\rangle = \left|\mathrm{BCS}^{(O,e,1)}\right\rangle, \quad E_0^{(O,e,1)} = |J+h| - (J+h) - \sum_{k\in\mathcal{K}^{(O,e,1)}} \omega(k) \stackrel{\substack{J>0 \\ h/J\to 0}}{=\!=\!=} -LJ \tag{S32}$$

where $k_*^{(L,N_F,a)} = \arg\min_{k\in\mathcal{K}^{(L,N_F,a)}, 0<k<\pi} \omega(k)$, so for $J>0$, e.g., $k_*^{(O,e,0)} = \pi/L$ if $h<0$ and $k_*^{(O,e,0)} = (L-2)\pi/L$ if $h>0$.

From the exact solutions above, we can see why $a=0$ for even $L$ and $a=1$ for odd $L$ are more natural choices. Such choices would consistently yield $-LJ$ as the ground state energy in the $h/J\to 0$ limit when $J>0$. With the choices of $a$ above, for both even $L$ and odd L, the ground state in the even-$N_F$ sector is the BCS-like state, and in the odd-$N_F$ sector is the BCS-like state with an extra $k=\pi$ excitation.

Furthermore, the even-$N_F$ sector and the odd-$N_F$ sector are both present in the same Hamiltonian, and the true ground state is given by the sector with a lower ground state energy. With $h>0$, we can find that

$$E_0^{(E,o,0)} - E_0^{(E,e,0)} = 2h - \sum_{k\in\mathcal{K}^{(E,o,0)}} \omega(k) + \sum_{k\in\mathcal{K}^{(E,e,0)}} \omega(k) > 0, \tag{S33}$$

$$E_0^{(O,o,1)} - E_0^{(O,e,1)} \geq - \sum_{k\in\mathcal{K}^{(O,o,1)}} \omega(k) + \sum_{k\in\mathcal{K}^{(O,e,1)}} \omega(k) > 0. \tag{S34}$$

Therefore, the even-$N_F$ sector (BCS-like ground states) have lower energies than the odd-$N_F$ sector. In other words, with our choice of boundary conditions for the Pauli operators (PBC for even $L$ and ABC for odd $L$) and positive $J$ and $h$, the ground state in the fermion representation is the BCS-like state which is physically natural.

It is also useful to give the position-space solutions in the small-$h/J$ limit when $J>0$, $h>0$. With the single-qubit eigenstates of $\tilde\sigma^x$ being $|\tilde\pm\rangle = (|\tilde\uparrow\rangle \pm |\tilde\downarrow\rangle)/\sqrt{2}$, the almost degenerate ground states in even-$N_F$ and odd-$N_F$ sectors are

$$\left|E_0^{(L,e,0)}\right\rangle = \left(\left|\tilde+\tilde-\tilde+\tilde-\cdots\right\rangle + (-1)^L \left|\tilde-\tilde+\tilde-\tilde+\cdots\right\rangle\right)/\sqrt{2} = \left(\left|\downarrow\uparrow\downarrow\uparrow\cdots\right\rangle + (-1)^L \left|\uparrow\downarrow\uparrow\downarrow\cdots\right\rangle\right)/\sqrt{2}, \tag{S35}$$

$$\left|E_0^{(L,o,0)}\right\rangle = \left(\left|\tilde+\tilde-\tilde+\tilde-\cdots\right\rangle - (-1)^L \left|\tilde-\tilde+\tilde-\tilde+\cdots\right\rangle\right)/\sqrt{2} = \left(\left|\downarrow\uparrow\downarrow\uparrow\cdots\right\rangle - (-1)^L \left|\uparrow\downarrow\uparrow\downarrow\cdots\right\rangle\right)/\sqrt{2}, \tag{S36}$$

where the states with $\tilde{\ }$ are in the eigenbasis of $\tilde\sigma^\alpha$ used for the discussion in this section, and the states without $\tilde{\ }$ are in the eigenbasis of $\sigma^\alpha$ in the original form of TFIM. The energy difference between the ground state and the first excited state, when even-$N_F$ and odd-$N_F$ sectors are considered in the same Hamiltonian, is $\Delta E = E_0^{(L,o,0)} - E_0^{(L,e,0)} \sim (h/J)^L$.

## III. THEORETICAL ANALYSIS OF LCD PARAMETER $\lambda_f$ FOR TFIM

For the adiabatic or counterdiabatic driving in our case, as the dynamical process with a time-dependent Hamiltonian is not exactly solvable at the general time $0 < t < \tau$, the exact analytic forms of state evolution cannot be obtained. In this section, for the choice of the LCD parameter $\lambda_f$ for TFIM, we provide theoretical analysis and intuition without exactly solving the dynamics.

For an arbitrary time-dependent unitary operator $U(t)$, the moving-frame transformation is

$$|\tilde{\psi}\rangle = U^\dagger |\psi\rangle \tag{S37}$$

Then the Schrödinger equation becomes

$$i\partial_t|\tilde{\psi}\rangle = iU^\dagger \partial_t |\psi\rangle + i\left(\partial_t U^\dagger\right)|\psi\rangle = \left(U^\dagger H U + i\left(\partial_t U^\dagger\right) U\right)|\tilde{\psi}\rangle \tag{S38}$$

Therefore, the moving frame Hamiltonian

$$\tilde{H}_m = U^\dagger H U + i\left(\partial_t U^\dagger\right) U. \tag{S39}$$

The unitary transformation does not have to diagonalize the instantaneous Hamiltonian $H$.

With the Ising model, we consider the $Y$-rotation $U_y(t) = \otimes_{i=1}^{L} R_y(\theta(t)) = e^{-i\theta_y(t) \sum_i \sigma_i^y/2}$ where $\theta_y(t) = 2\lambda_f \int_0^t dt'\, \dot{\lambda}(t')\alpha(t')$. Then in this rotating frame, the $\sigma^y$-terms can be cancelled exactly, resulting in the following effective Hamiltonian

$$\begin{aligned}
\widetilde{H}_{\text{LCD}}(t) &= U_y(t)^\dagger H_{\text{LCD}}(t) U_y(t) + i\left(\partial_t U_y(t)^\dagger\right) U(t) \\
&= [h_x(t)\cos\theta_y(t) - h_z(t)\sin\theta_y(t)]\sum_i \sigma_i^x + [h_z(t)\cos\theta_y(t) + h_x(t)\sin\theta_y(t)]\sum_i \sigma_i^z \\
&\quad + J(t)\sum_i \left(\sigma_i^z \cos\theta_y(t) - \sigma_i^x \sin\theta_y(t)\right) \otimes \left(\sigma_{i+1}^z \cos\theta_y(t) - \sigma_{i+1}^x \sin\theta_y(t)\right).
\end{aligned} \tag{S40}$$

As it is equivalent to work in the rotating frame physically, the transition amplitude to the target ground state (also transformed into the rotating frame) would be the same as that in the original frame

$$\begin{aligned}
F'(t) &= \langle \text{GS}|\mathcal{T}\exp\left(-i\int_0^t dt'\, H_{\text{LCD}}(t')\right)|\psi(0)\rangle \\
&= \left\langle \widetilde{\text{GS}}(\lambda_f)\left|\mathcal{T}\exp\left(-i\int_0^t dt'\, \widetilde{H}_{\text{LCD}}(t')\right)\right|\widetilde{\psi}(0)\right\rangle,
\end{aligned} \tag{S41}$$

where $\mathcal{T}$ is the time-ordering operator. Since $U_y(0)$ is the identity operator, we have $\left|\widetilde{\psi}(0)\right\rangle = |\psi(0)\rangle = |1\rangle^{\otimes L}$. The target TFIM Hamiltonian in this $Y$-rotating frame is not the same as the original frame:

$$\begin{aligned}
\widetilde{H}(\tau) &= \widetilde{H}_{\text{LCD}}(\tau) \\
&= h_{xf}\cos\theta_y(\tau)\sum_i \sigma_i^x + h_{xf}\sin\theta_y(\tau)\sum_i \sigma_i^z \\
&\quad + J_f \sum_i \left(\sigma_i^z \cos\theta_y(\tau) - \sigma_i^x \sin\theta_y(\tau)\right)\left(\sigma_{i+1}^z \cos\theta_y(\tau) - \sigma_{i+1}^x \sin\theta_y(\tau)\right),
\end{aligned} \tag{S42}$$

which has a frequency in the $\lambda_f$-space

$$\nu_{\lambda_f} = \frac{1}{\pi}\int_0^\tau dt'\, \dot{\lambda}(t')\alpha(t'), \tag{S43}$$

due to the $\lambda_f$-dependence of $\theta_y(\tau)$. The ground state $\left|\widetilde{\text{GS}}(\lambda_f)\right\rangle$ in the rotating frame, as a function of $\lambda_f$, is not the same as the one in the original frame $|\text{GS}\rangle$ but is transformed by the unitary $U_y(\tau)$. From Eq. S41 and Eq. S42,

the original LCD process is physically equivalent to the adiabatic process from the initial state to the $\lambda_f$-dependent $\left|\widetilde{\text{GS}}(\lambda_f)\right\rangle$.

The original TFIM Hamiltonian has a global symmetry $\otimes_i(-\sigma_i^x)$ (*i.e.*, the fermion parity) for any value of $h_{xf}$, and its ground state has the even fermion parity (with quantum number +1). In the rotating frame, at $t = \tau$, the symmetry operator becomes $\otimes_i(-\sigma_i^x \cos\theta(\tau) - \sigma_i^z \sin\theta(\tau))$ and can be checked to commute with $\tilde{H}(\tau)$. When we choose $\theta(\tau) = \pi/2 + 2\pi n$ for $n \in \mathbb{Z}$ in the LCD protocol, the target symmetry operator is $\otimes_i(-\sigma_i^z)$, and the initial state $\left|\widetilde{\psi}(0)\right\rangle = |1\rangle^{\otimes L}$ is parity-even with this symmetry. In other words, the initial state already has the same symmetry as $\left|\widetilde{\text{GS}}(\lambda_f)\right\rangle$ and can intuitively arrive at $\left|\widetilde{\text{GS}}(\lambda_f)\right\rangle$ through the adiabatic process more easily. We note that during the time evolution $0 < t < \tau$, the Hamiltonian $\tilde{H}_{\text{LCD}}(t)$ does not have the parity symmetry, and the argument above only considers the similarity between the initial and final states in the rotating frame. As we mention in the main text, in the limit of large $\lambda_f$, the rotating-frame time-dependent Hamiltonian changes very rapidly, and the fidelity starts to decay as shown in numerical simulations. We choose the $\lambda_f$ by setting $n = 0$ and thus $\theta(\tau) = \pi/2$. This is equivalent to our theoretical choice of $\lambda_{f,\text{opt}} = 1/(4\nu_{\lambda_f})$.

In the limit of large $h_{xf}$, the TFIM in the original frame gets reduced to a decoupled Hamiltonian with only $h_{xf} \sum_i \sigma_i^x$, and its ground state is $|-\rangle^{\otimes L}$ with $\sigma^x |\pm\rangle = \pm |\pm\rangle$. When the total evolution time $\tau$ is small, then the $X$ and $Z$ rotations from the time evolution of the Hamiltonian are not significant, and the $Y$ rotation from the LCD driving term dominates. In this case, the total effect of the LCD-term alone is a uniform $Y$-rotation with angle $\pi/2$ that brings the initial state $|1\rangle^{\otimes L}$ to $|-\rangle^{\otimes L}$.

## IV. HIGH-ORDER APPROXIMATE GAUGE POTENTIAL AND LU

### A. Expansion of the approximate gauge potentials

As is shown in Ref. [6], the exact gauge potential as an expansion

$$A_\lambda(\lambda) = \lim_{\epsilon \to 0^+} \int_0^\infty dt \, e^{-\epsilon t} \sum_{j=0}^\infty \frac{(-it)^j}{j!} \underbrace{[H, [H, \ldots [H, \partial_\lambda H] \cdots ]]}_{j} \tag{S44}$$

motivates the approximate potential to the $l$-th order as

$$A_\lambda^{(l)} = i \sum_{k=1}^l \alpha_k \underbrace{[H, [H, \ldots [H, \partial_\lambda H]]]}_{2k-1}. \tag{S45}$$

Here we consider the general form of the Hamiltonian in our procotol without counterdiabatic driving as

$$H(\lambda) = \sum_j \left( h_{xf} \lambda \sigma_j^x + J_f \lambda \sigma_j^z \sigma_{j+1}^z + h_{zi}(1-\lambda) \sigma_j^z \right), \tag{S46}$$

and thus

$$\partial_\lambda H(\lambda) = \sum_j \left( h_{xf} \sigma_j^x + J_f \sigma_j^z \sigma_{j+1}^z - h_{zi} \sigma_j^z \right). \tag{S47}$$

The first-order commutator corresponding to $k = 1$ is

$$[H, \partial_\lambda H] = 2i h_{xf} h_{zi} \sum_j \sigma_j^y, \tag{S48}$$

which is the form of the applied local CD driving in main text. The second-order commutator is not present in $A_\lambda$:

$$[H, [H, \partial_\lambda H]] = -4 h_{xf} h_{zi} \sum_j \left( h_{xf} \lambda \sigma_j^z - J_f \lambda \left( \sigma_j^x \sigma_{j+1}^z + \sigma_j^z \sigma_{j+1}^x \right) - h_{zi}(1-\lambda) \sigma_j^x \right). \tag{S49}$$

The third-order commutator corresponding to $k = 2$ is present in $A_\lambda$ and is calculated as

$$
\begin{aligned}
&[H, [H, [H, \partial_\lambda H]]] \\
&= -8ih_{xf}h_{zi} \sum_j \left[ -h_{xf}^2 \lambda^2 \sigma_j^y + h_{xf} J_f \lambda^2 \left( \sigma_j^x \sigma_{j+1}^y + \sigma_j^y \sigma_{j+1}^x \right) - 2J_f^2 \lambda^2 \sigma_j^z \sigma_{j+1}^y \sigma_{j+2}^z \right. \\
&\qquad \left. -2J_f h_{zi}\lambda(1-\lambda) \left( \sigma_j^y \sigma_{j+1}^z + \sigma_j^z \sigma_{j+1}^y \right) - h_{zi}^2 (1-\lambda)^2 \sigma_j^y \right].
\end{aligned}
\tag{S50}
$$

Similar to the Ising model in Ref. [7] (and Ref. [1] with a different protocol in the parameter $\lambda$), we consider the ansatz with up to two-body terms presented in Eq. S50

$$
A_\lambda^{(2)} = \alpha \sum_j \sigma_j^y + \gamma \sum_j \left( \sigma_j^x \sigma_{j+1}^y + \sigma_j^y \sigma_{j+1}^x \right) + \zeta \sum_j \left( \sigma_j^z \sigma_{j+1}^y + \sigma_j^y \sigma_{j+1}^z \right),
\tag{S51}
$$

which then gives

$$
\begin{aligned}
G_\lambda &= \partial_\lambda H + i[A_\lambda, H] \\
&= (-h_{zi} + 2\alpha h_x) \sum_j \sigma_j^z + (h_{xf} - 2\alpha h_z - 4\zeta J) \sum_j \sigma_j^x \\
&\quad + (J_f - 4\zeta h_x) \sum_j \sigma_j^z \sigma_{j+1}^z + (-2\alpha J + 2\gamma h_x - 2\zeta h_z) \sum_j \left( \sigma_j^x \sigma_{j+1}^z + \sigma_j^z \sigma_{j+1}^x \right) \\
&\quad + (4\gamma h_z - 4\zeta h_x) \sum_j \sigma_j^y \sigma_{j+1}^y - 4\gamma h_z \sum_j \sigma_j^x \sigma_{j+1}^x \\
&\quad + 2\gamma J \sum_j \left( -\sigma_j^x \sigma_{j+1}^x \sigma_{j+2}^z - \sigma_j^z \sigma_{j+1}^x \sigma_{j+2}^x + \sigma_j^x \sigma_{j+1}^y \sigma_{j+2}^y + \sigma_j^y \sigma_{j+1}^y \sigma_{j+2}^x \right) - 4\zeta J \sum_j \sigma_j^z \sigma_{j+1}^x \sigma_{j+2}^z.
\end{aligned}
\tag{S52}
$$

Then the action of the gauge potential is

$$
\begin{aligned}
&2^{-L} S[A_\lambda]/L \\
&= 2^{-L} \operatorname{Tr}\left( G_\lambda^2 \right)/L \\
&= (-h_{zi} + 2\alpha h_x)^2 + (h_{xf} - 2\alpha h_z - 4\zeta J)^2 + (J_f - 4\zeta h_x)^2 + 2(-2\alpha J + 2\gamma h_x - 2\zeta h_z)^2 \\
&\quad + (4\gamma h_z - 4\zeta h_x)^2 + (4\gamma h_z)^2 + 4(2\gamma J)^2 + (4\zeta J)^2.
\end{aligned}
\tag{S53}
$$

The action is just a quadratic function of $\alpha$, $\gamma$, and $\zeta$, so its minimization can be solved with the following linear equations

$$
2^{-L} \partial_\alpha \operatorname{Tr}\left( G_\lambda^2 \right)/L = 8\left( h_x^2 + h_z^2 + 2J^2 \right) \alpha - 16 h_x J \gamma + 32 h_z J \zeta - 4\left( h_{x,f} h_z + h_x h_{z,i} \right) = 0
\tag{S54}
$$

$$
2^{-L} \partial_\gamma \operatorname{Tr}\left( G_\lambda^2 \right)/L = -16 h_x J \alpha + 16\left( h_x^2 + 4h_z^2 + 2J^2 \right) \gamma - 48 h_x h_z \zeta = 0
\tag{S55}
$$

$$
2^{-L} \partial_\zeta \operatorname{Tr}\left( G_\lambda^2 \right)/L = 32 h_z J \alpha - 48 h_x h_z \gamma + 16\left( 4h_x^2 + h_z^2 + 4J^2 \right) \zeta - 8\left( J_f h_x + J h_{x,f} \right) = 0
\tag{S56}
$$

or equivalently, in the matrix form,

$$
8 \begin{pmatrix} (h_x^2 + h_z^2 + 2J^2) & -2h_x J & 4h_z J \\ -2h_x J & 2(h_x^2 + 4h_z^2 + 2J^2) & -6h_x h_z \\ 4h_z J & -6h_x h_z & 2(4h_x^2 + h_z^2 + 4J^2) \end{pmatrix} \begin{pmatrix} \alpha \\ \gamma \\ \zeta \end{pmatrix} = \begin{pmatrix} 4(h_{xf} h_z + h_x h_{zi}) \\ 0 \\ 8(J_f h_x + J h_{xf}) \end{pmatrix}.
\tag{S57}
$$

Substituting $h_x = h_{xf}\lambda$, $h_z = h_{zi}(1-\lambda)$, $J = J_f \lambda$, $h_{zi} = J_f$ and in the the small-$h_{xf}$ limit we have

$$
\alpha_{\text{LCD}}^{(2)} := \alpha = \frac{1 - 2\lambda - 3\lambda^2 + 8\lambda^3}{2(1 - 4\lambda + 4\lambda^2 + 7\lambda^4)} \frac{h_{x,f}}{J_f} + O\left( h_{xf}^3 \right),
\tag{S58}
$$

$$
\gamma = O\left( h_{xf}^2 \right),
\tag{S59}
$$

$$\zeta = \frac{-\lambda^2 + 3\lambda^3}{1 - 4\lambda + 4\lambda^2 + 7\lambda^4} \frac{h_{xf}}{J_f} + O\left(h_{xf}^3\right). \tag{S60}$$

In the other limit $h_{xf} \to \infty$,

$$\alpha^{(2)} := \alpha = \frac{J_f}{2\lambda^2 h_{xf}} + O\left(h_{xf}^{-3}\right) \tag{S61}$$

$$\gamma = \frac{5 - 3\lambda}{4\lambda^2} \frac{J_f^2}{h_{xf}^2} + O\left(h_{xf}^{-4}\right) \tag{S62}$$

$$\zeta = \frac{J_f}{4\lambda h_{xf}} + O\left(h_{xf}^{-3}\right), \tag{S63}$$

where $\alpha^{(2)}$ denotes the coefficient of $\sum_j \sigma_j^y$ with the $k=1$ and $k=2$ terms combined (not the same as $\alpha_{k=2}$ in Eq. S45). We highlight that in the small $h_{xf}$ limit, $\alpha$ and $\zeta$ have similar amplitudes. Note that in both the small and large $h_{xf}$ limits, $\gamma$ is much smaller than $\alpha$ and $\beta$ thus the interaction $\sigma_j^x \sigma_{j+1}^y + \sigma_j^y \sigma_{j+1}^x$ is less dominant than $\sum_j \sigma_j^y$ and $\sigma_j^z \sigma_{j+1}^y + \sigma_j^y \sigma_{j+1}^z$ in the second-order AGP with up to two-body terms.

As a comparison, in the first-order AGP (LCD for TFIM), in the small-$h_{xf}$ limit we have

$$\alpha^{(1)} = \frac{h_{xf} h_{zi}}{2\left(h_x^2 + h_z^2 + 2J^2\right)} = \frac{1}{2\left(1 - 2\lambda + 3\lambda^2\right)} \frac{h_{xf}}{J_f} + O\left(h_{xf}^3\right), \tag{S64}$$

and in the large-$h_{xf}$ limit, we have

$$\alpha^{(1)} = \frac{h_{xf} h_{zi}}{2\left(h_x^2 + h_z^2 + 2J^2\right)} = \frac{J_f}{2\lambda^2 h_{xf}} + O\left(h_{xf}^{-3}\right). \tag{S65}$$

In the large-$h_{xf}$ limit, we can see immediately that $\alpha^{(1)}$ and $\alpha^{(2)}$ agree at order $\mathcal{O}(1/h_{xf})$. In the small-$h_{xf}$ limit, at order $\mathcal{O}(h_{xf})$, to qualitatively compare the scales of $\alpha^{(1)}$ and $\alpha^{(2)}$, we compute the integrals over time of the $y$-amplitudes in the LCD Hamiltonians

$$\int_0^\tau dt' \, \dot\lambda \alpha^{(1)}(\lambda(t')) = \int_0^1 d\lambda \, \alpha^{(1)}(\lambda) = \frac{\pi}{4\sqrt{2}} \frac{h_{x,f}}{J_f} + O\left(h_{x,f}^3\right) = 0.5554 \frac{h_{x,f}}{J_f} + O\left(h_{x,f}^3\right), \tag{S66}$$

$$\int_0^\tau dt' \, \dot\lambda \alpha^{(2)}(\lambda(t')) = \int_0^1 d\lambda \, \alpha^{(2)}(\lambda) = 0.4304 \frac{h_{x,f}}{J_f} + O\left(h_{x,f}^3\right). \tag{S67}$$

The two integrals above differ by $\sim 30\%$ and are both independent of $\tau$ but dependent on $h_{xf}$.

### B. Fixed-angle LU and approximate gauge potentials

In this section, we illustrate the physical intuition of the fixed-angle LU. We conjecture that the local single-qubit rotations can effectively induce the two-body terms that show up in the second-order approximate gauge potential of the model, leading to better approximation of the exact AGP thus suppressing diabatic transitions.

We first introduce a corollary for interchanging the order between $e^A$ and $e^B$ where $A$ and $B$ matrices have low matrix norm but do not commute with each other. From the Baker-Campbell-Hausdorff (BCH) formula to the first order,

$$e^A e^B \approx e^{A+B+\frac{1}{2}[A,B]}. \tag{S68}$$

The Zassenhaus formula [8] to the first order is

$$e^{A+B} \approx e^A e^B e^{-\frac{1}{2}[A,B]}. \tag{S69}$$

A corollary from the two formula together is

$$\begin{aligned} e^A e^B &\approx e^{\left(B+\frac{1}{2}[A,B]\right)+A} \\ &\approx e^{B+\frac{1}{2}[A,B]} e^A e^{-\frac{1}{2}\left[B+\frac{1}{2}[A,B],A\right]} \\ &\approx e^{B+\frac{1}{2}[A,B]} e^A e^{\frac{1}{2}[A,B]} \\ &\approx e^{B+\frac{1}{2}[A,B]} e^{\frac{1}{2}[A,B]} e^A \\ &\approx e^{B+[A,B]} e^A. \end{aligned} \tag{S70}$$

To derive the third line, the second-order commutator $[[A,B],A]$ is neglected. To derive the fourth line, we use $\left[e^A, e^{\frac{1}{2}[A,B]}\right] \approx 0$, which holds true at order of the first-order commutators. For the derivation in the fifth line, the commutators at the second and higher orders are neglected.

Eq. S70 provides the formula to move $e^A$ from the left to the right of $e^B$, while introducing an additional term $[A,B]$ to the exponential of $B$. We next apply this corollary to the post-LCD LU rotation. The Trotterized expression for time evolution under the LCD Hamiltonian is

$$\begin{aligned} &U_{\text{LCD}}(\tau,0) \\ &= \lim_{N\to\infty} \exp\left(-i\Delta t\, H_{\text{LCD}}(t_N)\right) \exp\left(-i\Delta t\, H_{\text{LCD}}(t_{N-1})\right) \cdots \exp\left(-i\Delta t\, H_{\text{LCD}}(0)\right), \end{aligned} \tag{S71}$$

where $t_k = k\tau/N$ and $\Delta t = \tau/N$. Therefore,

$$\begin{aligned} &\exp\left(-i\frac{\theta}{2}\sum_i \sigma_i^x\right) U_{\text{LCD}}(t,0) \\ &\approx \exp\left(-i\frac{\theta}{2}\sum_i \sigma_i^x\right) \exp\left(-i\Delta t\, H_{\text{LCD}}(t_N)\right) \exp\left(-i\Delta t\, H_{\text{LCD}}(t_{N-1})\right) \cdots \exp\left(-i\Delta t\, H_{\text{LCD}}(0)\right) \\ &= \exp\left(-i\Delta t\, H_{\text{LCD}}(t_N) - \frac{\theta}{2}\Delta t\left[\sum_i \sigma_i^x, H_{\text{LCD}}(t_N)\right]\right) \exp\left(-i\frac{\theta}{2}\sum_i \sigma_i^x\right) \\ &\quad \exp\left(-i\Delta t\, H_{\text{LCD}}(t_{N-1})\right)\cdots \exp\left(-i\Delta t\, H_{\text{LCD}}(0)\right) \\ &= \exp\left(-i\Delta t\, H_{\text{LCD}}(t_N) - \frac{\theta}{2}\Delta t\left[\sum_i \sigma_i^x, H_{\text{LCD}}(t_N)\right]\right) \exp\left(-i\Delta t\, H_{\text{LCD}}(t_{N-1}) - \frac{\theta}{2}\Delta t\left[\sum_i \sigma_i^x, H_{\text{LCD}}(t_{N-1})\right]\right) \\ &\quad \exp\left(-i\frac{\theta}{2}\sum_i \sigma_i^x\right) \exp\left(-i\Delta t\, H_{\text{LCD}}(t_{N-2})\right) \cdots \exp\left(-i\Delta t\, H_{\text{LCD}}(0)\right) \\ &= \exp\left(-i\Delta t\, H_{\text{LCD}}(t_N) - \frac{\theta}{2}\Delta t\left[\sum_i \sigma_i^x, H_{\text{LCD}}(t_N)\right]\right) \exp\left(-i\Delta t\, H_{\text{LCD}}(t_{N-1}) - \frac{\theta}{2}\Delta t\left[\sum_i \sigma_i^x, H_{\text{LCD}}(t_{N-1})\right]\right) \\ &\quad \cdots \exp\left(-i\Delta t\, H_{\text{LCD}}(0) - \frac{\theta}{2}\Delta t\left[\sum_i \sigma_i^x, H_{\text{LCD}}(0)\right]\right) \exp\left(-i\frac{\theta}{2}\sum_i \sigma_i^x\right). \end{aligned} \tag{S72}$$

The effective Hamiltonian considering the $X$-rotation can be read from Eq. S72 as

$$\begin{aligned} H_{\text{LCD}}^{\text{eff}}(t) &= H_{\text{LCD}}(t) - i\frac{\theta}{2}\left[\sum_i \sigma_i^x, H_{\text{LCD}}(t)\right] \\ &= H_{\text{LCD}}(t) - i\frac{\theta}{2}\sum_{i,j}\left[\sigma_i^x, h_x(t)\sigma_j^x + h_y(t)\sigma_j^y + h_z(t)\sigma_j^z + J(t)\sigma_j^z\sigma_{j+1}^z\right] \\ &= H_{\text{LCD}}(t) + \theta\sum_i \left(h_y(t)\sigma_i^z - h_z(t)\sigma_i^y - J(t)\left(\sigma_i^y\sigma_{i+1}^z + \sigma_i^z\sigma_{i+1}^y\right)\right). \end{aligned} \tag{S73}$$

The local unitary $X$-rotation after the LCD acting on the initial state $|\psi(0)\rangle$ can be viewed as

$$\exp\left(-i\frac{\theta}{2}\sum_i \sigma_i^x\right) U_{\text{LCD}}(\tau,0)|\psi(0)\rangle$$
$$\approx U_{\text{LCD}}^{\text{eff}}(\tau,0)\exp\left(-i\frac{\theta}{2}\sum_i \sigma_i^x\right)|\psi(0)\rangle, \quad (S74)$$

where $U_{\text{LCD}}^{\text{eff}}(\tau,0)$ is the time evolution operator of $H_{\text{LCD}}^{\text{eff}}(t)$.

Eq. S73 indicates that the $X$-rotation effectively introduces additional $\sigma_i^y \sigma_{i+1}^z + \sigma_i^z \sigma_{i+1}^y$ interactions that are also present in the second-order approximate gauge potential in Sec. IV A. In addition, they are more dominant than the other two-body term in the small and large $h_{xf}$ limits, as shown above. The $\theta \sum_i h_y(t)\sigma_i^z$ term vanishes at initial and final time which might play a similar role as the $\sigma_i^z$-terms used in Ref. [7, 9].

We note that the initial and final Hamiltonians are modified by the LU. At the initial time $t=0$, since we have $h_y(0)=0$ for the LCD $Y$-term and $J(0)=0$, $h_z(0)=h_{zi}=1$, the initial effective LCD Hamiltonian is

$$H_{\text{LCD}}^{\text{eff}}(0) = H_{\text{LCD}}(0) - \theta \sum_i \sigma_i^y = \sum_i (\sigma_i^z - \theta \sigma_i^y) \quad (S75)$$

with the ground state being the product state $(i\frac{(1-\sqrt{1+\theta^2})}{\theta}|0\rangle + |1\rangle)^{\otimes L}$ where the normalization factor is omitted. Recall that $|\psi(0)\rangle = |1\rangle^{\otimes L}$ and thus $\exp(-i\theta\sum_i \sigma_i^x/2)|\psi(0)\rangle = (-i\sin(\theta/2)|0\rangle + \cos(\theta/2)|1\rangle)^{\otimes L}$, which, to the linear order of $\theta$, is the ground state of $H_{\text{LCD}}^{\text{eff}}(0)$. Therefore, in the effective picture above, the $X$-rotated state as the new initial state is approximately the ground state of $H_{\text{LCD}}^{\text{eff}}(0)$. This reconciles the additional term in the initial effective LCD Hamiltonian and the $X$-rotation directly on $|\psi(0)\rangle$.

We note that the discussion above to derive the effective LCD Hamiltonian induced by the $X$-rotation is heuristic and not rigorous, since the derivation for $H_{\text{LCD}}^{\text{eff}}(t)$ might require $\theta$ being small. In addition, the $\theta J(t)(\sigma_i^y \sigma_{i+1}^z + \sigma_i^z \sigma_{i+1}^y)$ interaction induced by LU rotation to mimic the second-order approximate gauge potential would also modify the target Hamiltonian $H_{\text{LCD}}^{\text{eff}}(\tau)$ by order $\mathcal{O}(\theta)$, since $J(\tau) \neq 0$. This may lead to additional discrepancy by order $\mathcal{O}(\theta)$ of the final point in the path of the counterdiabatic process.

## V. ADDITIONAL SIMULATION RESULTS ON PROTOCOL PERFORMANCE

### A. LCD performance at small and intermediate $h_{xf}$ regime

In Fig. 1 of the manuscript we focus on the performance of LCD driving at $h_{xf} = 2$. The LCD protocol has a much worse performance in the regime of small $h_{xf}$. In Fig. S1, we show the results at $h_{xf} = 0.5$, using $L=4$ as an example. We further show the scaling behavior of fidelity at $h_{xf} = 1$ in Fig. S2.

### B. Robustness against noise for LCD driving

Introducing noise channels during the LCD evolution inevitably results in decreased fidelity. To understand the impact of noise, we conduct numerical simulations on state fidelity using various strengths of depolarization ($\kappa_1$) and dephasing ($\kappa_2$) noise, as illustrated in Fig. S3. The system's density matrix is determined by solving the Lindblad master equation.

$$\dot{\rho}(t) = -i[H_{\text{LCD}}(t), \rho(t)] + \sum_n \frac{1}{2}(2C_n \rho(t) C_n^\dagger - \rho(t) C_n^\dagger C_n - C_n^\dagger C_n \rho(t)), \quad (S76)$$

where $C_n = \sqrt{\kappa_{1(2)}} A_n$ specifies the collapse operators, and $A_n$ are the decay operators (we consider $X(Z)$ operators for depolarization (dephasing) channel) and $\kappa_{1(2)}$ are the corresponding rates. As expected, the fidelity drops exponentially for both noise types. For the depolarization channel analyzed, the fidelities of LCD-based protocols converge to $1/2^L$ with the adiabatic approach, while they diminish to zero for the dephasing channel.

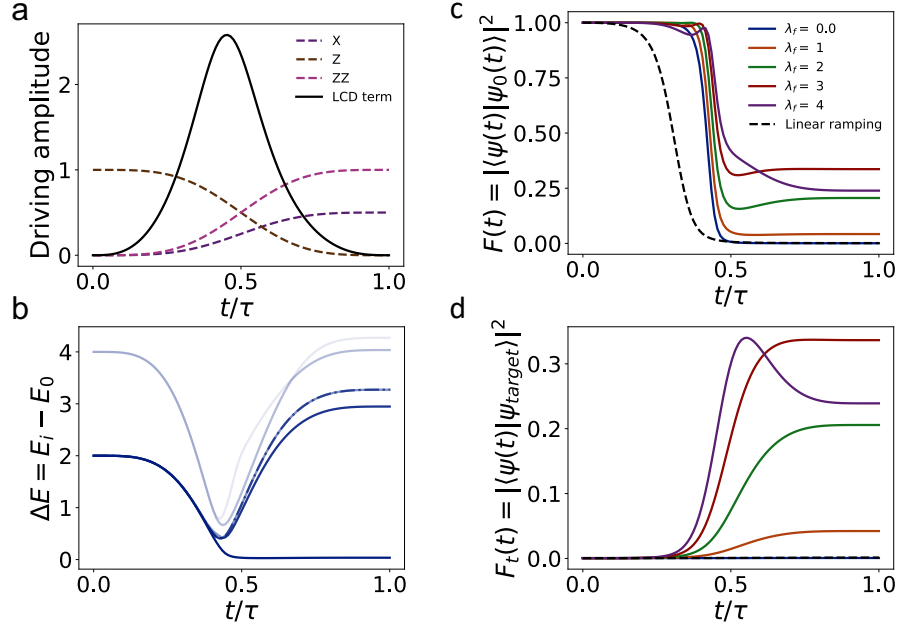

FIG. S1. Dynamics and performance of LCD driving for $h_{xf} = 0.5$. **a.** Time evolution of the driving amplitudes. LCD term corresponds to local $Y$ operators with $\lambda_f = 3.095$. **b.** Instantaneous energy spectrum of $H(t)$ for $h_{xf} = 0.5$. Energy difference between the first six excited states and the ground state is shown. Here the third and fourth excited states are degenerate. Note that TFIM has degenerate ground states when $h_{xf} < 1$ in thermodynamic limit. **c-d.** Fidelity of the system state with respect to the instantaneous ground state of $H_0(t)$ (c) and target state to be prepared (d) for $h_{xf} = 0.5$. The $\lambda_f = 0$ curve is the adiabatic protocol using the sweep function specified in main text, while the linear ramping curve corresponds to linear adiabatic control $H_{\text{linear}}(t) = (1 - t/\tau) \sum_{i=1}^{L} \sigma_i^z + t/\tau (\sum_{i=1}^{L} h_{xf} \sigma_i^x + \sum_{i=1}^{L-1} \sigma_i^z \sigma_{i+1}^z)$. The system size is $L = 4$ for the plots shown here.

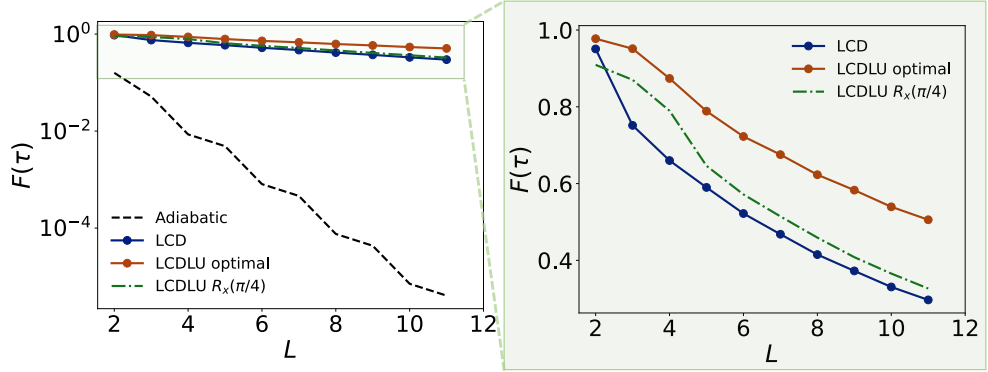

FIG. S2. Fidelity as a function of system size $L$ up to 12 qubits at fixed $h_{xf} = 1.0$. The right plot provides an enlarged view of the results.

### C. Simulation results for varying total time $\tau$

Until now, our analysis has focused on evolution over a fixed period $J_f \tau = 1$. Extending the evolution time within the LCD protocol, as anticipated, leads to improved fidelity for LCD, albeit requiring significantly more time for the adiabatic approach to exhibit increased fidelity (Fig. S4a). Particularly, when $J_f \tau > 10$, the protocol can achieve a near-unity fidelity due to the fact the slow sweeping function leads to a much smaller diabatic transition rate. Conversely, the fixed-angle LCDLU approach, which utilizes an $X$-rotation for the LU, exhibits superior performance relative to LCD when $J_f \tau \approx 1$, with the optimal LU corresponding to $X$-rotations where the rotation angle is approximately $\pi/4$. In the scenario of large $J_f \tau$, given that LCD already approximates the target state closely, the optimal LU angle approaches zero. (Fig. S4b).

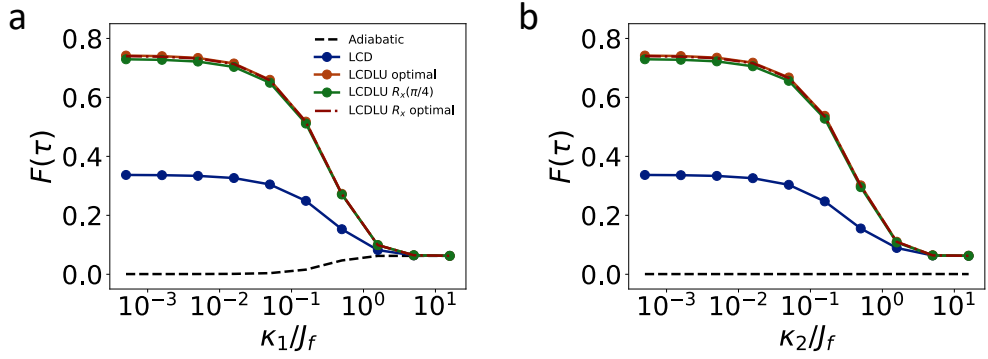

FIG. S3. Fidelity as a function of depolarization strength $\kappa_1$ (a) and dephasing strength $\kappa_2$ (b) at fixed $h_{xf} = 0.5$ using $L = 4$ as an example.

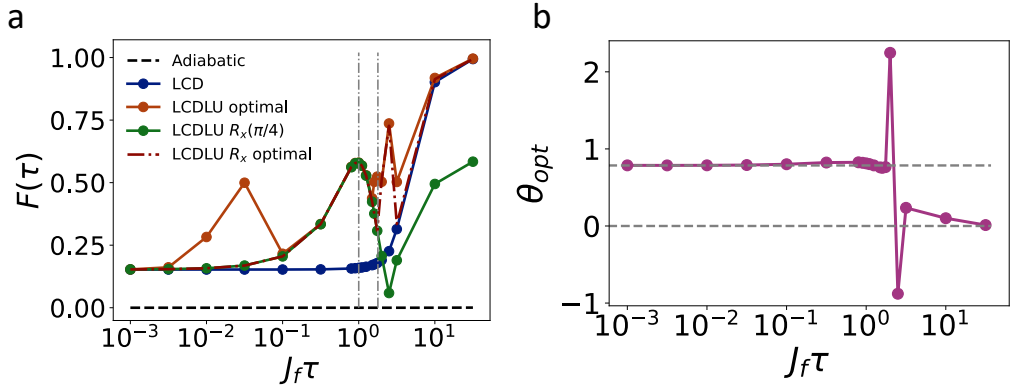

FIG. S4. **a.** Fidelity of prepared states as a function of LCD evolution time $\tau$ at fixed $h_{xf} = 0.1$ and $L = 4$. The dashed grey lines mark the point $J_f\tau = 1$ and $J_f\tau = 1 + \pi/4$. **b.** Optimal $X$-rotation angle for LU found by numerical simulations. The horizontal and vertical grey lines in the figure correspond to $\pi/4(0)$ and $J_f\tau = 1$, respectively. The optimal LCDLU in (a) at $J_f\tau \sim 0.03$ has a fidelity around 0.5, while the numerical optimization at lower $J_f\tau$ fails to find such high fidelity. The nonuniform optimal LCDLU with a short total time $J_f\tau < 0.1$ can only prepare a product state $|0101\rangle$ or $|1010\rangle$ when the numerical optimization in LU converge to the global maximum of fidelity. At $J_f\tau \lesssim 0.01$, our numerical optimization for nonuniform LCDLU in (a) turns out to be inefficient as it is in the overparameterized parameter space with vanishing gradient. We note that this regime of small $\tau$ and small $h_{xf}$ is not practically interesting for LCD.

We further remark that when selecting $J_f\tau = 1$ for LCD, as described in the main text, the total evolution time of LCDLU would be $(1 + \frac{\pi}{4})\tau$ if we consider the LU process to be an evolution under X driving with fixed amplitude:

$$\text{LU} = \exp\left(-i\frac{\pi}{4}\sum_i \frac{\sigma_i^x}{2}\right) = \exp\left(-i\int_0^{\frac{\pi}{4}} \sum_i \frac{\sigma_i^x}{2} dt\right). \tag{S77}$$

As illustrated in Fig. S4a, the performance of LCDLU is considerably better than that of LCD for the same evolution duration.

The regime of small $\tau$ and small $h_{xf}$ is not practically interesting for LCD, since the effect of counterdiabatic or adiabatic evolution in this case is small. From the perspective of numerical optimization, in the $3L$-dimensional parameter space of $\{\alpha_i, \beta_i, \gamma_i\}_{i=1}^L$, when the total LCD time $\tau$ is very small, the prepared state from LCDLU is close to product state, thus lying in the $2L$-dimensional product-state submanifold of the $2^L$-dimensional Hilbert space. Now the $3L$ real parameters from the arbitrary local unitary rotation overparameterize the product-state submanifold. The $3L$-dimensional numerical optimization thus has vanishing gradient in a relatively large parameter space, so the optimizer may fail to find the true global maximum, as the points for LCDLU optimal in Fig. S4a at $J_f\tau \lesssim 0.01$ show.

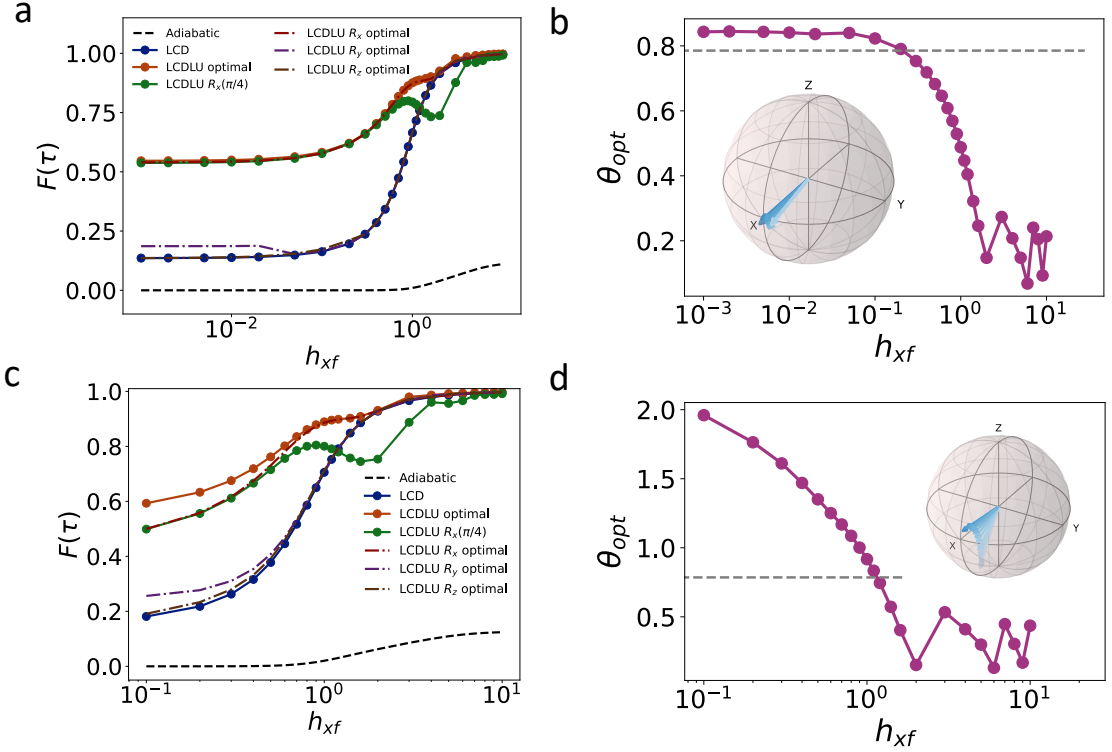

FIG. S5. Performance of LCD and LCDLU protocols when $h_{zf} \neq 0$ in the target Hamiltonian. **a-b.** $h_{zf} = 0.1$. **c-d.** $h_{zf} = 0.5$. The figures on the right (b,d) shows the optimal angle and axis for the global single-body rotation in LU. The grey dashed line corresponds to $\pi/4$. The inset Bloch spheres show the rotation axis for $h_{xf} \lesssim 1$, where smaller $h_{xf}$ corresponds to more transparent vectors. Here the evolution time $\tau$ is fixed to be 1.

### D. Simulation results for nonintegrable model

To showcase the performance of the LCD-based protocols for models that are not exactly solvable, we consider the task of preparing ground state of

$$H(\tau) = \sum_{i=1}^{L} h_{zf}\sigma_i^z + \sum_{i=1}^{L} h_{xf}\sigma_i^x + \sum_{i=1}^{L} J_f \sigma_i^z \sigma_{i+1}^z \tag{S78}$$

with $h_{zf} \neq 0$. The initial state is still the ground state of $\sum_i^L \sigma_i^z$ as other simulations. As shown in Fig. S5, we present the results of fidelity change for various $h_{xf}$ at fixed $h_{zf} = 0.1(0.5)$.

In the scenario of $\sqrt{h_{xf}^2 + h_{zf}^2} \gg J_f = 1$, the final Hamiltonian is dominated by the single-body terms, and the $Y$-rotation angle needed during the LCD evolution from the initial $Z$ direction to the direction defined by $h_{xf}\sum_i \sigma_i^x + h_{zf}\sum_i \sigma_i^z$ is given by $\arctan(h_{xf}/h_{zf})$. Similar as previous cases, the LCD achieves a high fidelity in the single-body limit and the optimal LU rotation angle is close to zero.

In general, when $h_{zf} \neq 0$, the model becomes an nonintegrable Ising model that is not exactly solvable. For LCD, the operator in this case is still $\sum_i \sigma_i^y$, our theoretical prediction for the optimal $\lambda_f$ still holds. For LCDLU, we show that in the nonintegrable case, the site-independent uniform LU $LU(\alpha, \beta, \theta) = (R_z(\alpha)R_x(\theta)R_z(\beta))^{\otimes L}$ yields about the same best fidelity compared to the optimal LU in the most general form. This indicates that the translationally symmetric LU is already an optimal choice for LCDLU in this example.

We re-parameterize the uniform LU as $\bigotimes_i \exp(-i\theta\mathbf{n}\cdot\boldsymbol{\sigma}_i/2)$ and further show the rotational axis given by the direcitonal vector $\mathbf{n} = (n_x, n_y, n_z)$ and rotational angle $\theta$. When $h_{zf} = 0.1$ (Fig. S5a-b), the optimal LU axis is still $X$ for varying $h_{zf}$ and the results in main text hold here; while when $h_{zf} = 0.5$ (Fig. S5c-d), the optimal rotational axis deviates from the $X$-rotation, and the rotational angle is different from $\theta = \pi/4$. In the latter case, the target Hamiltonian differs significantly from the exactly solvable TFIM and the model does not preserve the parity

symmetry $\otimes_i(-\sigma_i^x)$. Therefore, contrary to the $h_{zf} = 0$ case studied in main text, the optimal LU operation would not be $X$-rotation. Indeed, when $h_{xf} \to 0$, the optimal rotation axis would start to go from $X$ to the $-Z$ direction.

## VI. EXPERIMENTAL DETAILS

### A. Experimental setup

The experiments in this work were performed on a trapped-ion system (Quantinuum H1-1 platform) [10]. The qubit information is encoded in the hyperfine approximate clock states of $^{171}$Yb$^+$ in the $^2S_{1/2}$ state. Namely, $|0\rangle \equiv |F = 0, m_f = 0\rangle$ and $|1\rangle \equiv |F = 1, m_f = 0\rangle$. After loading, the qubits can be prepared in $|0\rangle$ via optical pumping [10, 11].

The system design is based on the QCCD architecture with multiple (five for H1-1) separate gate zones with a single linear geometry [10]. Each gate zone is used to perform operations on an arbitrary pair of two qubits at a time, suppressing crosstalk and maintaining high fidelity. Although in the experiments we study models with nearest-neighbour interactions, the system has all-to-all connectivity with two-qubit gates between pairs of qubits implemented by ion transport, which brings the pairs into the same gate zone, allowing us to implement periodic boundary conditions.

To implement two-qubit gates, a phase-sensitive Mølmer-Sørensen (MS) gate sandwiched between single-qubit wrapper pulses is used. It in turn gives the $ZZ$ gate $R_{ZZ}(\gamma) = \exp\left(-i\gamma \frac{ZZ}{2}\right)$, where the rotation angle can be precisely controlled by changing the parameters in the MS gate [10, 11]. The H1-1 system used in this work has a typical average two-qubit infidelity of $2 \times 10^{-3}$, with single qubit infidelity two orders of magnitude smaller ($\approx 4 \times 10^{-5}$). The qubit state can be read out via state-dependent resonance fluorescence measurement. Note that mid-circuit measurement and reset can be implemented while causing a small crosstalk error due to the stray light from the measurement and reset laser beams. The state preparation and measurement error rate is about $3 \times 10^{-3}$.

We note that the highest error operation on the devices is the two-qubit gate, so the primary limiting factor for achieving high-fidelity results is the two-qubit gate count. For the Trotterized circuit that we implement in this work, the number of two-qubit gates for implementing nearest-neighbour $ZZ$ interactions is given by $N_{2q} = LT_{steps}$ where $T_{steps}$ is the number of Trotter steps.

### B. Measurement of observables

As shown in Fig.3 of the main text, in each Trotter step we implement $H_{\text{LCD}}$ with the rotation angles determined by the LCD schedule. In the quantum state tomography (QST) experiment, we perform measurement on the $3^L$ Pauli strings $\{X, Y, Z\}^L$ by applying corresponding single-body rotation gate right before the measurement. Therefore, for the $L = 4$ QST experiment, 81 circuits are run to construct density matrix.

To extract the average energy of the evolved states of LCD and LCDLU, we measure the state in the $Z$ and $X$ bases to extract

$$|\langle H_0(\tau)\rangle| = \left|\sum_i (h_{xf}\langle \sigma_i^x\rangle + \langle \sigma_i^z \sigma_{i+1}^z\rangle)\right|.$$

Note that as the LU only invovles $X$-rotation gates, we have $\langle \sum_i \sigma_i^x \rangle_{\text{LCD}} = \langle \sum_i \sigma_i^x \rangle_{\text{LCDLU}}$. For the experiment results, we conduct standard error propagation, utilizing measurement errors in $\sum_i \langle \sigma_i^x \rangle$ and $\sum_i \langle \sigma_i^z \sigma_{i+1}^z \rangle$ to compute the error in $|\langle H_0(\tau)\rangle|$, assuming maximum covariance between the two measurements.

We observe that the expected energy $\langle H_0(\tau)\rangle$, which is straightforward to measure experimentally, can serve as a reliable metric for benchmarking the fidelity of prepared states relative to the target ground state for small $L$, as it's shown in Fig. 2a and Fig. 3d in the main text. This observation holds only for the small system sizes. However, in generic large system sizes, even though the system remains gapped (except when $h_{xf} \neq 1$), minor perturbations in the control parameters could lead to orthogonal states (known as "orthogonality catastrophe") [12]. In such cases, the average energy no longer represents the fidelity of the prepared states.

We remark that the energy fluctuation could be another metric to benchmark the fidelity of prepared states,

$$\sigma^2(H_0(\tau)) = \langle H_0(\tau)^2\rangle - \langle H_0(\tau)\rangle^2.$$

However, there would be $O(L^2)$ non-commuting observables required to construct $\langle H_0^2(\tau)\rangle$.

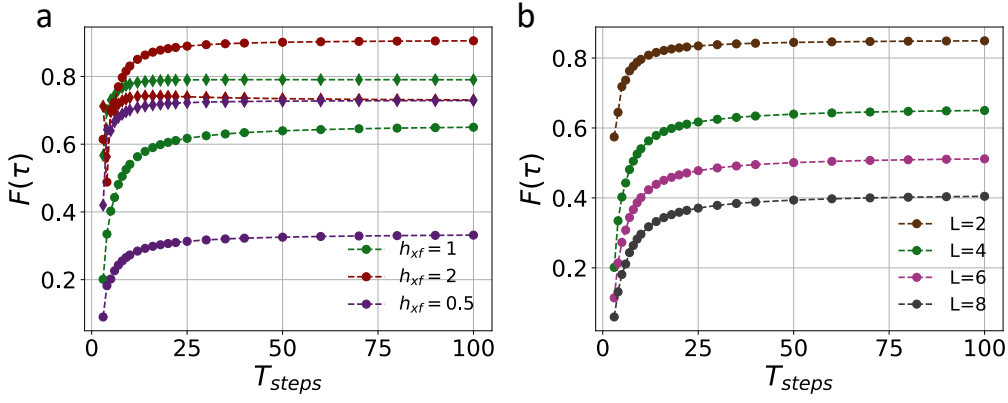

FIG. S6. Noiseless simulation on state fidelity as a function of Trotter steps in the circuit for fixed $L = 4$ (a) and fixed $h_{xf} = 1$ (b). The circles (diamonds) show the data points for state prepared by LCD (LCDLU). The LU rotation here is fixed $R_x$ rotation with with an angle of $\pi/4$.

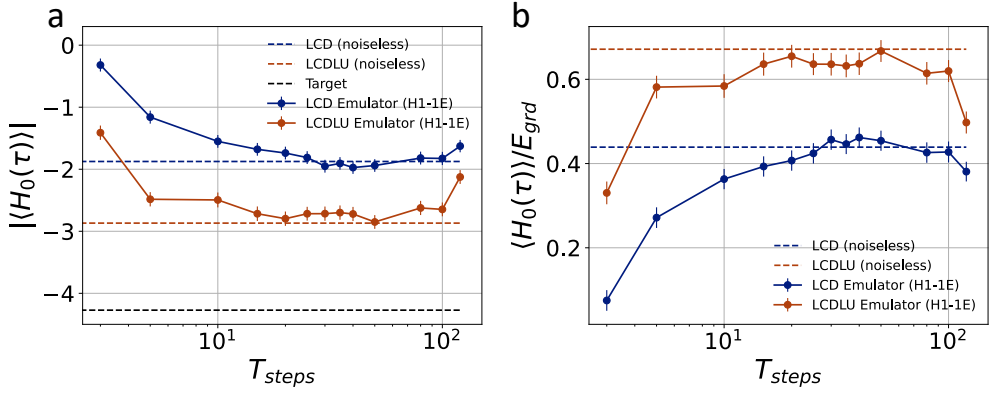

FIG. S7. Noisy simulations on the average energy of the state prepared by LCD and LCDLU. The simulations are performed in the Quantinuum H1-1E emulator that uses the same noise model and parameters as the H1-1 hardware device. **a.** Average energy **b.** Ratio of average energy with respect to the target ground state energy. The data shown is for $L = 4$ and $h_{xf} = 0.5$. The LU rotation here is fixed $R_x$ rotation with an angle of $\pi/4$.

### C. Trotterization of LCD evolution

To identify the optimal number of Trotter steps, we conduct numerical simulations on state fidelity as a function of $T_{steps}$. The noiseless simulation depicted in Fig. S6 reveals that the fidelity of the state prepared by LCD driving saturates at $T_{steps} \approx 20$. Furthermore, thanks to the high two-qubit fidelity in the trapped-ion system, performance does not begin to degrade until $T_{steps} \approx 100$ for systems containing 4 qubits, as shown in the average energy plots in Fig. S7. This translates into a total number of two-qubit gate count around 400. In the experiments conducted for this work, we utilized a fixed number of Trotter steps set to 20 in order to minimize losses due to gate errors while still achieving performance comparable to that of continuous evolution.

### D. Additional experimental data and analysis

In this section, we provide further experimental data to complement the results outlined in the main text. In Fig. S8, we present the imaginary part of the density matrix constructed from QST experiments. We note reduced magnitudes for the imaginary components as a consequence of the applied LU evolution. As a comparison, density matrix of the target state has no imaginary entries.

In Fig. S9, we show the measured probability distribution of bitstrings in the Z-basis for system size $L = 10$

17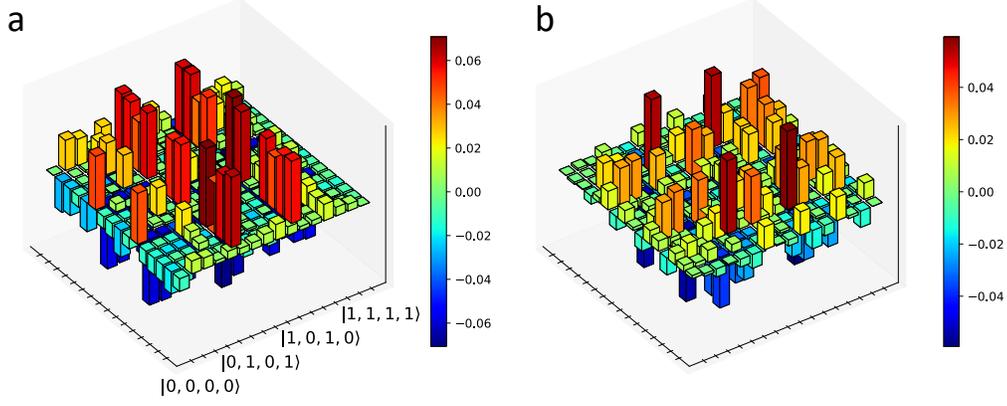

FIG. S8. Imaginary part of the density matrix constructed from quantum state tomography for LCD state (a) and LCDLU state (b). The system contains four qubits and $h_{xf}$ is set to be 0.5 here. The target density matrix has no imaginary entries.

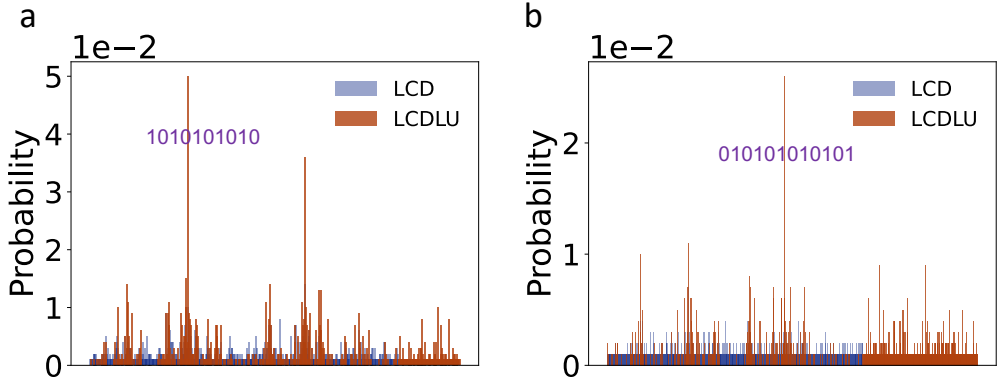

FIG. S9. Probability distribution of bitstrings obtained from measurements in the Z-basis on states prepared by LCD and LCDLU. $L = 10$ (a) and $L = 12$ (b). The number of measurements is 1000 for both plots. The bitstring with largest probability is highlighted in purple text.

and $L = 12$. Even with diminishing fidelity as the system size expands, the LCDLU technique consistently produces bitstrings exhibiting antiferromagnetic ordering with high probability. The number of circuit repetitions is 1000 for all the $Z(X)$ basis measurements when extracting expected energy, except that the number of shots for $Z$ measurement when $L = 14$ is 850.

The evidence of antiferromagnetic ordering in the target ground state in the small $h_{xf}$ regime is further supported by its two-point correlations. To characterize the performance of LCD and LCDLU, in Fig. S10, we present the site correlations $|\langle Z_i Z_j \rangle|$ for states produced by LCD and LCDLU at $h_{xf} = 0.5$, both in noiseless simulation and noisy experiments. Notably, LCDLU demonstrates enhanced ZZ correlations compared to LCD, though these correlations decrease to zero with increasing site distance. The noiseless simulation yields a fidelity around 0.17 for the state prepared by the fixed-angle LCDLU in this 10-qubit system.

[1] M. Kolodrubetz, D. Sels, P. Mehta, and A. Polkovnikov, Geometry and non-adiabatic response in quantum and classical systems, Physics Reports **697**, 1 (2017), geometry and non-adiabatic response in quantum and classical systems.
[2] D. Sels and A. Polkovnikov, Minimizing irreversible losses in quantum systems by local counterdiabatic driving, Proceedings of the National Academy of Sciences **114**, 10.1073/pnas.1619826114 (2017).
[3] P. W. Claeys, M. Pandey, D. Sels, and A. Polkovnikov, Floquet-engineering counterdiabatic protocols in quantum many-body systems, Phys. Rev. Lett. **123**, 090602 (2019).
[4] G. E. Santoro, Non-equilibrium quantum systems, International School for Advanced Studies Lecture Notes , 1 (2016).

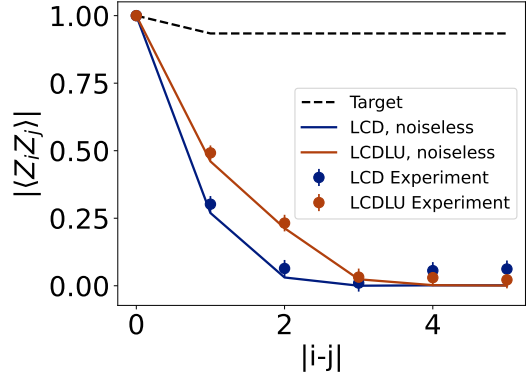

FIG. S10. Example of distance-dependent average site correlations measured on $L = 10$ system. Here the transverse field strength of target Hamiltonian is $h_{xf} = 0.5$. The values for $\langle Z_i Z_j \rangle$ are positive (negative) when $|i - j|$ is even (odd) for both simulation and experiment results.



\end{document}